\documentclass[twocolumn,tighten]{aastex701}

\usepackage{colortbl,comment}
\usepackage{xcolor}
\usepackage{xspace}
\usepackage[version=4]{mhchem}

\newcommand{\e}{\mbox{e}^}
\newcommand{\amm}{\ce{NH_3}\xspace}
\newcommand{\ccs}{\ce{C2S}\xspace}
\newcommand{\hcfn}{\ce{HC_5N}\xspace}
\newcommand{\hcsn}{\ce{HC_7N}\xspace}

\newcommand{\dia}{\ce{N_2H^+}\xspace}

\newcommand{\namm}{\ensuremath{N(\amm)}\xspace}
\newcommand{\xamm}{\ensuremath{X(\amm)}\xspace}

\newcommand{\nhcfn}{\ensuremath{N(\ce{HC_5N})}\xspace}

\newcommand{\sqcm}{\ensuremath{\mathrm{cm}^{-2}}\xspace}

\newcommand{\vlsr}{\ensuremath{v_{\mathrm{LSR}}}\xspace}
\newcommand{\sigv}{\ensuremath{\sigma_v}\xspace}
\newcommand{\signt}{\ensuremath{\sigma_\mathrm{NT}}\xspace}
\newcommand{\csound}{\ensuremath{c_\mathrm{s}}\xspace}
\newcommand{\tkin}{\ensuremath{T_\mathrm{K}}\xspace}
\newcommand{\tex}{\ensuremath{T_\mathrm{ex}}\xspace}

\newcommand{\avir}{\ensuremath{\alpha_\mathrm{vir}}\xspace}

\newcommand{\msun}{\ensuremath{\rm M_\odot}\xspace}

\newcommand{\DRone}{GAS DR1\xspace}

\newcommand{\kms}{\ensuremath{\mathrm{km\,s}^{-1}}\xspace}
\newcommand{\ms}{\ensuremath{\mathrm{m\,s}^{-1}}\xspace}
\newcommand{\nh}{\ensuremath{N(\ce{H2})}\xspace}

\newcommand\T{\rule{0pt}{2.6ex}} 
\newcommand\B{\rule[-1.2ex]{0pt}{0pt}} 

\shorttitle{GAS: Data Release 2}
\shortauthors{Pineda \& Friesen et al.}
\submitjournal{ApJS}
\accepted{2025 Sept. 05}

\begin{document}

\title{The Green Bank Ammonia Survey: 
Data Release 2}

\correspondingauthor{J. E. Pineda}
\email{jpineda@mpe.mpg.de}

\author[0000-0002-3972-1978]{Jaime E. Pineda}
\email{jpineda@mpe.mpg.de}
\affiliation{Max Planck Institute for Extraterrestrial Physics, Gie{\ss}enbachstra{\ss}se 1, D-85748 Garching bei M{\"u}nchen, Germany}

\author[0000-0001-7594-8128]{Rachel K. Friesen}
\email{rachel.friesen@utoronto.ca}
\affiliation{Department of Astronomy \& Astrophysics, University of Toronto, 50 St. George St., Toronto, ON, M5S 3H4, Canada}

\collaboration{2}{(co-PIs)}

\author[0000-0002-5204-2259]{Erik Rosolowsky}
\email{rosolowsky@ualberta.ca}
\affiliation{Department of Physics, University of Alberta, Edmonton, AB, Canada} 

\author[0000-0001-8804-8604]{Ana Chac\'on-Tanarro}
\email{a.chacon.tanarro@gmail.com}
\affiliation{Observatorio Astron\'omico Nacional (OAN-IGN), Alfonso XII 3, E-28014 Madrid, Spain}

\author[0000-0003-4242-973X]{Michael Chun-Yuan Chen}
\email{chen.m@queensu.ca}
\affiliation{Department for Physics, Engineering Physics and Astrophysics, Queen's University, Kingston, ON, K7L 3N6, Canada}

\author[0000-0002-9289-2450]{James Di Francesco}
\email{James.DiFrancesco@nrc-cnrc.gc.ca}
\affiliation{Department of Physics and Astronomy, University of Victoria, 3800 Finnerty Road, Victoria, BC, Canada V8P 5C2}
\affiliation{Herzberg Astronomy and Astrophysics, National Research Council of Canada, 5071 West Saanich Road, Victoria, BC, V9E 2E7, Canada}

\author[0000-0002-5779-8549]{Helen Kirk}
\email{helenkirkastro@gmail.com}
\affiliation{Department of Physics and Astronomy, University of Victoria, 3800 Finnerty Road, Victoria, BC, Canada V8P 5C2}
\affiliation{Herzberg Astronomy and Astrophysics, National Research Council of Canada, 5071 West Saanich Road, Victoria, BC, V9E 2E7, Canada}

\author[0000-0001-6004-875X]{Anna Punanova}
\email{punanovaanna@gmail.com}
\affiliation{Onsala Space Observatory, Chalmers University of Technology, Observatoriev\"agen 90, R\aa\"o, 439 92 Onsala, Sweden}

\author[0000-0003-2122-2617]{Youngmin Seo}
\email{seo3919@gmail.com}
\affiliation{Steward Observatory, 933 North Cherry Avenue, Tucson, AZ 85721, USA}

\author[0000-0002-0133-8973]{Yancy Shirley}
\email{yancyshirley@gmail.com}
\affiliation{Steward Observatory, 933 North Cherry Avenue, Tucson, AZ 85721, USA}

\author[0000-0001-6431-9633]{Adam Ginsburg}
\email{adam.g.ginsburg@gmail.com}
\affiliation{Department of Astronomy, University of Florida, P.O. Box 112055, Gainesville, FL 32611-2055, USA}

\author[0000-0003-1252-9916]{Stella S. R. Offner}
\email{offner@gmail.com}
\affiliation{Department of Astronomy, The University of Texas, Austin, TX 78712, USA}

\author[0000-0002-8897-1973]{Ayush Pandhi}
\email{ayush.pandhi@mail.utoronto.ca}
\affiliation{Department of Astronomy \& Astrophysics, University of Toronto, 50 St. George St., Toronto, ON, M5S 3H4, Canada}

\author[0000-0001-6022-3618]{Ayushi Singh}
\email{ayushi.singh@mail.utoronto.ca}
\affiliation{Sidrat Research, 124 Merton Street, Suite 507, Toronto, Ontario, M4S 2Z2, Canada}

\author[0009-0006-9160-1021]{Feiyu Quan}
\email{feiyu.quan@mail.utoronto.ca}
\affiliation{Department of Applied Mathematics and Theoretical Physics, University of Cambridge, Cambridge, CB3 0WA, UK}

\author[0000-0001-5653-7817]{H\'ector G. Arce}
\email{hector.arce@yale.edu}
\affiliation{Department of Astronomy, Yale University, P.O. Box 208101, New Haven, CT 06520-8101, USA}

\author[0000-0003-1481-7911]{Paola Caselli}
\email{caselli@mpe.mpg.de}
\affiliation{Max Planck Institute for Extraterrestrial Physics, Gie{\ss}enbachstra{\ss}se 1, D-85748 Garching bei M{\"u}nchen, Germany}

\author[0000-0002-7497-2713]{Spandan Choudhury}
\email{spandan@kasi.re.kr}
\affiliation{Max Planck Institute for Extraterrestrial Physics, Gie{\ss}enbachstra{\ss}se 1, D-85748 Garching bei M{\"u}nchen, Germany}
\affiliation{Korea Astronomy and Space Science Institute, 776 Daedeok-daero Yuseong-gu, Daejeon, 34055, Republic of Korea}

\author[0000-0003-1312-0477]{Alyssa A. Goodman}
\email{agoodman@cfa.harvard.edu}
\affiliation{Harvard-Smithsonian Center for Astrophysics, 60 Garden St., Cambridge, MA 02138, USA}

\author[0000-0002-4775-039X]{Fabian Heitsch}
\email{fheitsch@email.unc.edu}
\affiliation{Department of Physics and Astronomy, University of North Carolina Chapel Hill, Chapel Hill, NC 27599, USA}

\author[0000-0002-5236-3896]{Peter G. Martin}
\email{pgmartin@cita.utoronto.ca}
\affiliation{Canadian Institute for Theoretical Astrophysics, University of Toronto, 60 St. George St., Toronto, Ontario, Canada, M5S 3H8}

\author[0000-0001-9732-2281]{Christopher D. Matzner}
\email{matzner@astro.utoronto.ca}
\affiliation{Department of Astronomy \& Astrophysics, University of Toronto, 50 St. George St., Toronto, ON, M5S 3H4, Canada}

\author[0000-0002-2885-1806]{Philip C. Myers}
\email{pmyers@cfa.harvard.edu}
\affiliation{Harvard-Smithsonian Center for Astrophysics, 60 Garden St., Cambridge, MA 02138, USA}

\author[0000-0002-0528-8125]{Elena Redaelli}
\email{Elena.Redaelli@eso.org}
\affiliation{European Southern Observatory, Karl-Schwarzschild-Stra{\ss}e 2, 85748 Garching, Germany}
\affiliation{Max Planck Institute for Extraterrestrial Physics, Gie{\ss}enbachstra{\ss}se 1, D-85748 Garching bei M{\"u}nchen, Germany}

\author[0000-0002-9485-4394]{Samantha Scibelli}
\email{sscibell@nrao.edu}
\altaffiliation{Jansky Fellow of the National Radio Astronomy Observatory}
\affiliation{National Radio Astronomy Observatory, 520 Edgemont Road, Charlottesville, VA 22903, USA}

\collaboration{23}{(The GAS collaboration)}

\begin{abstract}
We present an overview of the final data release (DR2) from the Green Bank Ammonia Survey (GAS). 
GAS is a Large Program at the Green Bank Telescope to map all Gould Belt star-forming regions with $A_\mathrm{V} \gtrsim 7$~mag visible from the northern hemisphere in emission from \amm\ and other key molecular tracers. 
This final release includes the data for all the regions observed: Heiles Cloud 2 and B18 in Taurus; Barnard 1, Barnard 1-E, IC348, NGC 1333, L1448, L1451, and Per7/34 in Perseus; L1688 and L1689 in Ophiuchus; Orion~A (North and South) and Orion~B in Orion;
Cepheus, B59 in Pipe; Corona Australis (CrA) East and West; IC5146; and Serpens Aquila and MWC297 in Serpens. 
Similar to what was presented in GAS DR1, we find that the \amm\ emission and dust continuum emission from \textit{Herschel} correspond closely. 
We find that the \amm emission is generally extended beyond the typical 0.1\ pc length scales of dense cores, and we find that the transition between coherent core and turbulent cloud is a common result.
{This shows that the regions of coherence are common throughout different star forming regions, 
with a substantial fraction of the high column density regions displaying subsonic non-thermal velocity dispersions.}
We produce maps of the gas kinematics, temperature, and \amm\ column densities through forward modeling of the hyperfine structure of the \amm (1,1) and (2,2) lines. 
We show that the \amm\ velocity dispersion, \sigv, and gas kinetic temperature, \tkin, vary systematically between the regions included in this release, with an increase in both the mean value and spread of \sigv\ and \tkin\ with increasing star formation activity. 
The data presented in this paper are publicly available via \dataset[DOI: 10.11570/24.0091]{https://doi.org/10.11570/24.0091}.
\end{abstract}

\keywords{stars:formation --- ISM:molecules} 

\section{Introduction}
Stars form in molecular cores, the densest parts of molecular clouds, which provide the initial conditions for star formation \citep{Pineda2023-PP7}. 
Key properties providing information about core evolution are density, temperature, velocity, and degree of turbulence. 
Dense cores can be efficiently identified in dust continuum emission, since it is a good tracer of the 
total gas column density. 
Though such observations have resulted in catalogues of dense cores, they do not probe the gas kinematics.

Observations of dense cores in molecular lines are extremely useful, since they provide complementary information about the chemistry and kinematics of cores. 
Typical dense core tracers like \amm  and \dia are commonly used, since they hardly deplete from the gas phase \citep{Crapsi2007-L1544_VLA_Temperature,Redaelli2019-L1544_Deuteration,Caselli2022-L1544_NH2D_Depletion,Pineda2022-HMM1_NH3_Depletion,Lin2023-NH3_Depletion_Cores}. 
On the other hand, tracers like \ce{C^{18}O} or \ccs  are useful to study a different environment than the dense gas tracers lines. 
Chemical models predict that species like \ccs, \hcfn, and \hcsn have a peak in abundance at earlier times than species like \amm  or \dia \citep{Suzuki1992-CCS_NH3,Aikawa2001-Core_Chemistry}, 
and carbon-bearing species are more affected by molecular depletion \citep{Caselli1999-L1544_Depletion,Tafalla2002-Cores_Depletion}. 
It is for these two reasons that the two groups of molecules appear to trace different volumes.

Thanks to its meta-stable levels, \amm can serve as an efficient thermometer
\citep{Walmsley1983-NH3_Thermometer,HoTownes1983-NH3_Review}. 
In addition, the hyperfine structures of the \amm  lines enable a precise study of the kinematic information. 
Hence, \amm is widely used to determine dense core properties {\citep{Myers1983-Dense_Cores_NH3,Bachiller1986-Perseus_NH3,Benson1989-Dense_Cores_Survey,Jijina1999-NH3_Cores_Database,Tafalla+04,Wu2006-NH3_High_Mass,Rosolowsky2008-GBT_NH3_Perseus,Pineda2010-B5_Transition_to_Coherence,Purcell2012-HOPS_NH3_Catalogue,Wienen2012-NH3_ATLASGAL,Seo2015-Taurus_Core_Evolution,Feher2016-TMC1_Herschel_NH3}}.

The Gould Belt clouds include the nearby clouds {($<$1 kpc)}, and they have been studied in great detail with different observatories. 
The protostellar content is well determined with {\it Spitzer} observations \citep{Dunham2015-YSO_c2d}, the {\it Herschel} observations provide 
maps of the total column density, \nh, and dust temperature, while ground-based observations with SCUBA2 at the JCMT provide the 
dense core catalogs.

The initial data release, \DRone,  provided fully reduced data cubes for all lines observed, \amm (1,1),  (2,2), and (3,3), \ccs, \hcfn, and \hcsn, 
as well as the derived parameters of fitting the \amm (1,1) and (2,2) lines  for a limited number of regions  \citep{Friesen2017-GAS_DR1}.
In this second release, we now include all regions observed, as well as the derived parameters for the carbon-bearing species (\ccs, \hcfn, and \hcsn).
All data and the corresponding derived parameters are delivered as FITS files via CANFAR\footnote{\dataset[DOI: 10.11570/24.0091]{https://doi.org/10.11570/24.0091}}.

GAS \DRone\ data have been used to examine the structure and kinematics of dense gas within the nearby star-forming regions. 
An incomplete list of findings includes the following: 
(a) The smallest cores identified in GAS tend not to be bound by gravity but are instead pressure confined \citep{Kirk2017-OrionA_Pressure,Kerr2019-GAS_Pressure,Keown_GAS_2017}, while larger molecular clumps appear virialized \citep{Singh_2021_subvirial}; 
(b) Multiple velocity components in the \amm line profiles trace supersonic gas around subsonic cores in L1688 \citep{Choudhury_2020_supersonic}, and the quiescent core gas extends further out than previously argued \citep{Choudhury2021-L1688_Extended}; 
(c) Velocity gradients within cores and filaments suggest mass accretion onto and along filaments \citep{M_Chen+20};
(d) Core shapes and velocity gradients are generally randomly oriented with respect to the large-scale magnetic field traced by Planck \citep{CChen_PlanckGAS_2020,Pandhi2023-GAS_Shape_VGrad}. 

In this paper, we present the full GAS dataset. In Section~\ref{sec:data}, we describe the observations, calibration and gridding of the data. 
In Section~\ref{sec:analysis}, we describe the line fitting, moment maps, and column density calculations for all observed lines. 
In Section~\ref{sec:disc}, we discuss the bulk variation of some of the dense gas properties, and examine the use of carbon-chain vs. \amm emission to identify and analyse mass accretion via streamers in Perseus B1. We summarize our results in Section \ref{sec:summary} and provide links to the final dataset.

\section{Observations and data}
\label{sec:data}

\begin{figure*}
    \includegraphics[width=\textwidth]{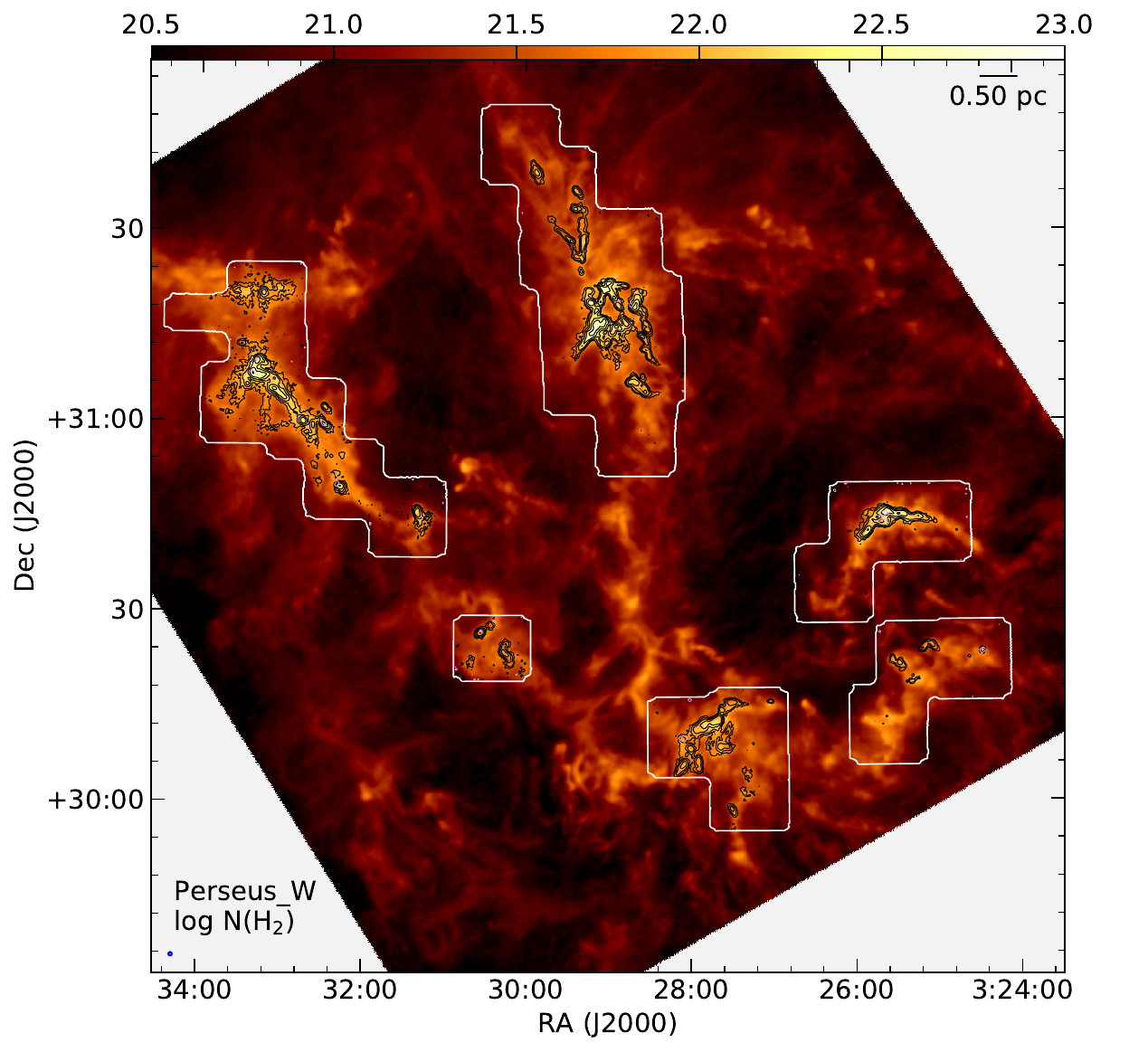}
    \caption{Observed GAS regions in the western Perseus molecular cloud. The color scale shows \nh 
    derived from SED fitting of Herschel continuum maps \citep{Singh2022-Tau_Herschel}. 
    White contours highlight the extent of the GAS maps. Black contours show the integrated 
    \amm\ (1,1) emission. The 32\arcsec\ GBT beam is shown at bottom left. \label{fig:perseus_w}}
\end{figure*}

The GAS survey goals were to map all regions {with visual extinctions} $A_\mathrm{V} {\gtrsim} 7$ mag toward the northern hemisphere-visible molecular clouds in the Gould Belt. 
The value of $A_\mathrm{V} \sim 7$ mag is motivated by a possible column density threshold for dense cores \citep[e.g.,][]{Johnstone2004-Oph_clump,Andre+10}, and by detectability of \amm in previous observations.
This selection comprises many of the star-forming clouds within $\sim 500$~pc of the Sun, covering a range in cloud sizes, masses, and star formation activity. 
Table \ref{tab:sources} lists the regions observed along with their distances and area mapped per region. Overall, GAS mapped $\sim 4$ deg$^2$ on sky over 9 clouds with distances ranging from $\sim 120$~pc to $\sim 400$~pc, with one cloud (IC 5146) at a distance of $\sim 800$~pc. 
An example of the coverage in Perseus is shown in Figure~\ref{fig:perseus_w}, which includes 
Barnard 1, Barnard 1E, NGC1333, Per7/34, L1448, and L1451.

Observations for the GAS survey were performed from January 2015 through March 2016 at the Robert C. Byrd Green Bank Telescope (GBT) using the 7-pixel K-band Focal Plane Array (KFPA) and the VErsatile GBT Astronomical Spectrometer (VEGAS). Observational details were first presented in \DRone, which focused on the \amm\ (1,1) and (2,2) emission from four regions. We refer the reader to \DRone\ \citep{Friesen2017-GAS_DR1} for a full description of the observations, data reduction, and imaging of the survey data. Here we discuss primarily observations of the additional spectral lines and targeted regions, as well as improvements to the GAS data reduction pipelines that are utilized in the data release associated with this paper. 

The VEGAS backend was used in its configuration Mode 20, allowing eight spectral windows per KFPA beam, each with a bandwidth of 23.44~MHz and 4096 spectral channels. The resulting spectral resolution of 5.7~kHz gives a velocity resolution of $\sim 0.07$~\kms at 23.7~GHz. The GAS setup used six of eight available spectral windows to target six spectral lines in each beam: the \amm\ (1,1) through (3,3) inversion transitions, plus the \hcfn\ and \hcsn\ rotational transitions listed in Table \ref{tab:params}. In addition, the \ce{C2S} $2_1-1_0$ transition (also in Table \ref{tab:params}) was observed in a single, central beam. In-band frequency switching with a frequency throw of 4.11~MHz was used, maximizing the observing time spent on-source. 

Observations were done in On-The-Fly (OTF) mode, most often scanning in right ascension (R.A.) or declination (decl.) over square regions of size 10$\arcmin\ \times$10$\arcmin$. The observed coverage of most regions consists of multiple such observing blocks to cover the desired area, generally observed on different dates. Most 10$\arcmin\times$ 10$\arcmin$ blocks were observed once to reach the survey sensitivity goals, but several were observed twice to mitigate the effects of poor weather in the first observations. In some regions, maps of 5$\arcmin\times$10$\arcmin$ were better suited to match the expected emission structure. Toward a few regions, the maps (and scan direction) were tilted relative to the R.A. and decl. axes. 
In OTF mode, scans were separated by 13\arcsec\ perpendicular to the scan direction to ensure Nyquist sampling (the beam is $\approx$32\arcsec at these frequencies). 
The telescope scan rate was 6.2\arcsec~s$^{-1}$, with data dumped every 1.044~s. Each 10$\arcmin\ \times$10$\arcmin$ map was completed in $1.4$~hours. Pointing updates were usually performed in between completed maps, or more frequently if winds were high ($> 5$~\ms). 

Calibration was performed using observations of the Moon and Jupiter, following the description in \DRone. Individual beam gains were determined per semester and applied to the data through the GAS calibration pipeline. The calibration pipeline is essentially the same as described in \DRone, with improved flagging of scans with outlying $T_{sys}$ values that improved the overall noise properties of the maps. The imaging pipeline is as described in \DRone. All GAS data reduction and imaging codes are publicly available on GitHub\footnote{\url{https://GitHub.com/GBTAmmoniaSurvey/GAS}}. 

The noise properties of the data will be discussed further in Appendix~\ref{appendix:rms}, as we used the results of the line fitting to refine the windows where the root mean square (rms) noise properties were calculated. 
We note here, however, that the \ce{C2S} emission maps have a factor $\sim \sqrt{7}$ higher rms noise, along with a smaller map footprint relative to lines observed with all seven KFPA beams due to being observed in a single beam only. In addition, the spectral window targeting the \hcsn\ (22--21) line often showed greater noise values than the others. 
Lastly, the spacing of the KFPA beams results in increased rms noise values at the map edges for the other lines. 
We masked the final data cubes by performing a binary erosion with a 2-dimensional disk of  three pixels diameter on the combined map, 
in addition to removing the {images'} edge prior to any further analysis.  

\begin{deluxetable}{lccccc}
  \tablecolumns{6}
  \tabletypesize{\footnotesize} 
  \tablewidth{0pt}
  \tablecaption{Observed regions, distances, and masses\label{tab:sources}}
  \tablehead{
    \colhead{Cloud} & \colhead{Region} & \colhead{Distance} & \colhead{Area} & \colhead{Mass}  & \colhead{Ref.} \\
    \colhead{} & \colhead{} & \colhead{(pc)} & \colhead{(\arcdeg$^2$)} & 
    \colhead{(\msun)} &
  }
  \startdata
  Taurus & HC2 & 138.6 & 0.35 & 271 & 1\\ 
  Taurus & B18 & 126.6 & 0.33 & 209 & 1\\ 
  Perseus & L1451 & 279 & 0.08 & 123 & 2\\ 
  Perseus & L1448 & 288 & 0.08 & 162 & 2\\
  Perseus & L1455 & 279 & 0.08 & 177 & 2\\
  Perseus & NGC1333 & 299 & 0.25 & 590  & 2\\
  Perseus & B1 & 301 & 0.15 & 363  & 2\\
  Perseus & B1E & 301 & 0.03 & 42  & 2\\
  Perseus & IC348 & 321 & 0.14 & 329  & 3\\ 
  Perseus & Per7/34 & 301 & 0.03 & 38  & 2 \\ 
  Ophiuchus & L1688 & 138.4 & 0.38 & 312   & 4 \\
  Ophiuchus & L1689 & 144.2 & 0.11 & 107  & 4 \\ 
  Ophiuchus & L1712 & 144.2 & 0.03 & 16   & 4 \\  
  OrionA & Orion~A & 397 & 0.39 & 2916  & 5 \\ 
  OrionA & Orion~A-S & 428 & 0.19 & 1995  & 5  \\
  OrionB & NGC 2023 & 403 & 0.17 & 778  & 6 \\
  OrionB & NGC 2068 & 417 & 0.12 & 90 & 6 \\
  IC 5146 &  & 813 & 0.17 & 389 & 7 \\
  CrA & CrA East & 154 & 0.03 & 8 & 7 \\
  CrA & CrA West & 154 & 0.06 & 25 & 7 \\
  Pipe & B59 & 163 & 0.03 & 22 & 7 \\
  Pipe & Core 40 & 163 & 0.01 & 59 & \\ 
  Serpens & Serpens Aquila & 436 & 0.35 & 3141  & 3 \\
  Serpens & MWC 297 & 436 & 0.03 & 262  & 3 \\
  Cepheus & L1228 & 346 & 0.10 & 228 \\ 
  Cepheus & L1251 & 346 & 0.24 & 473 & 8 \\
  \enddata
  \tablecomments{The mass listed is the total mass within the extent of the area mapped by GAS, 
  calculated from dust continuum opacity and temperatures maps derived from SED fitting of 
  submillimeter continuum data at similar resolution to the \amm\ observations \citep{Singh2022-Tau_Herschel}.}
\tablerefs{
  (1) \cite{galli_2018}, 
  (2) \cite{zucker_2018},
  (3) \cite{ortiz_leon_2018},
  (4) \cite{ortiz_leon_2018b}, 
  (5) \cite{Grosschedl2018-OrionA_GAIA_DR2},
  (6) \cite{Kounkel2018-Orion_APOGEE2_Distance}, 
  (7) \cite{dzib_2018},
  (8) \cite{yan_2019}
}
\end{deluxetable} 

\begin{deluxetable}{lcccc}
\T
\tablewidth{0pt}
\tablecolumns{5}
\tablecaption{Line parameters \label{tab:params}}
\tablehead{
\colhead{Species} & \colhead{Transition} & \colhead{$\nu$} & \colhead{$B_0$} & \colhead{$\mu$} \\
\colhead{} & \colhead{} & \colhead{(GHz)} & \colhead{(MHz)} & \colhead{(Debye)}}
\startdata
\hcfn & $J = 9-8$ & {23.963901} & 1331.33 & 4.33 \\
\hcsn & $J = 21-20$ & {23.6878974} & 564.0 & 4.82 \\
\hcsn & $J = 22-21$ & {24.8158772} & 564.0 & 4.82 \\ 
\ccs & $J_N = 2_1-1_0$ & {22.344030} & 6477.75 & 2.88 
\enddata
\end{deluxetable} 

\section{Analysis}
\label{sec:analysis}

The calibration and imaging pipelines described above produce data cubes in position-position-velocity (PPV) space. In regions larger than the typical 10$\arcmin\ \times$10$\arcmin$ observational footprint, map blocks are combined to cover contiguous areas toward each observed region. Table \ref{tab:sources} lists the total map area observed for each region; regions with areas greater than 0.03\arcdeg$^2$ contain multiple blocks. 

\subsection{Line fitting}

\subsubsection{Hyperfine model fitting of \amm (1,1) and (2,2)}

Following the method described in \DRone, we fit simultaneously the \amm\ (1,1) and (2,2) lines using the {\texttt cold\_ammonia} model implemented in \texttt{pyspeckit} \citep{pyspeckit2011,pyspeckit2022}. 
From the simultaneous fits, we obtain five fit parameters: \vlsr, \sigv, \tkin, \tex, and \namm. 
{We note that \tkin is derived assuming the filling factor is unity and the relation between rotational temperature and kinetic temperature derived by \cite{Swift2005-PreProtostellar_Core_L1551_NH3}, which assumes that both {transitions} have the same excitation temperature and velocity dispersion.}
We follow the masking rules adopted in \DRone \citep[see][for details]{Friesen2017-GAS_DR1} to only keep fitted parameters with a reliable determination. 

\subsubsection{Single Gaussian fitting of \amm (3,3) and carbon bearing species \ce{CCS}, \hcfn, and \hcsn}

The \amm ($J,K$) transitions observed here are inversion transitions, where $J$ is the total angular momentum quantum number, and $K$ is the component of $J$ along the molecular axis. \amm (3,3) is an ortho-\amm transition ($K = 3n$, where $n = 0,1,2,\ldots$), whereas \amm (1,1) and (2,2) are transitions of para-\amm ({$K \ne 3n$}). The para- and ortho-\amm states are not well-connected in typical conditions in dense clouds, and we thus treat \amm (3,3) as a separate species. While the line is also an inversion transition and contains hyperfine structure as the \amm (1,1) and (2,2) transitions, we do not detect it with sufficient signal-to-noise ratio (S/N) to detect the separate components, and consequently fit the line with a single Gaussian. 

For \amm (3,3) and each carbon-bearing species and transition, we fit all spectra with S/N $>3$ with a single Gaussian component, producing maps of line amplitude, $T_B$, line-of-sight velocity in the local standard of rest frame, \vlsr, and Gaussian line width, \sigv, along with their respective uncertainties. The parameter maps were masked where the uncertainty in \vlsr was greater than 0.3~\kms, and where \sigv was less than three times its fit uncertainty. 
Lastly, we masked pixels without neighbouring good fits. 

\begin{figure*}
    \centering
    \includegraphics[width=0.9\textwidth]{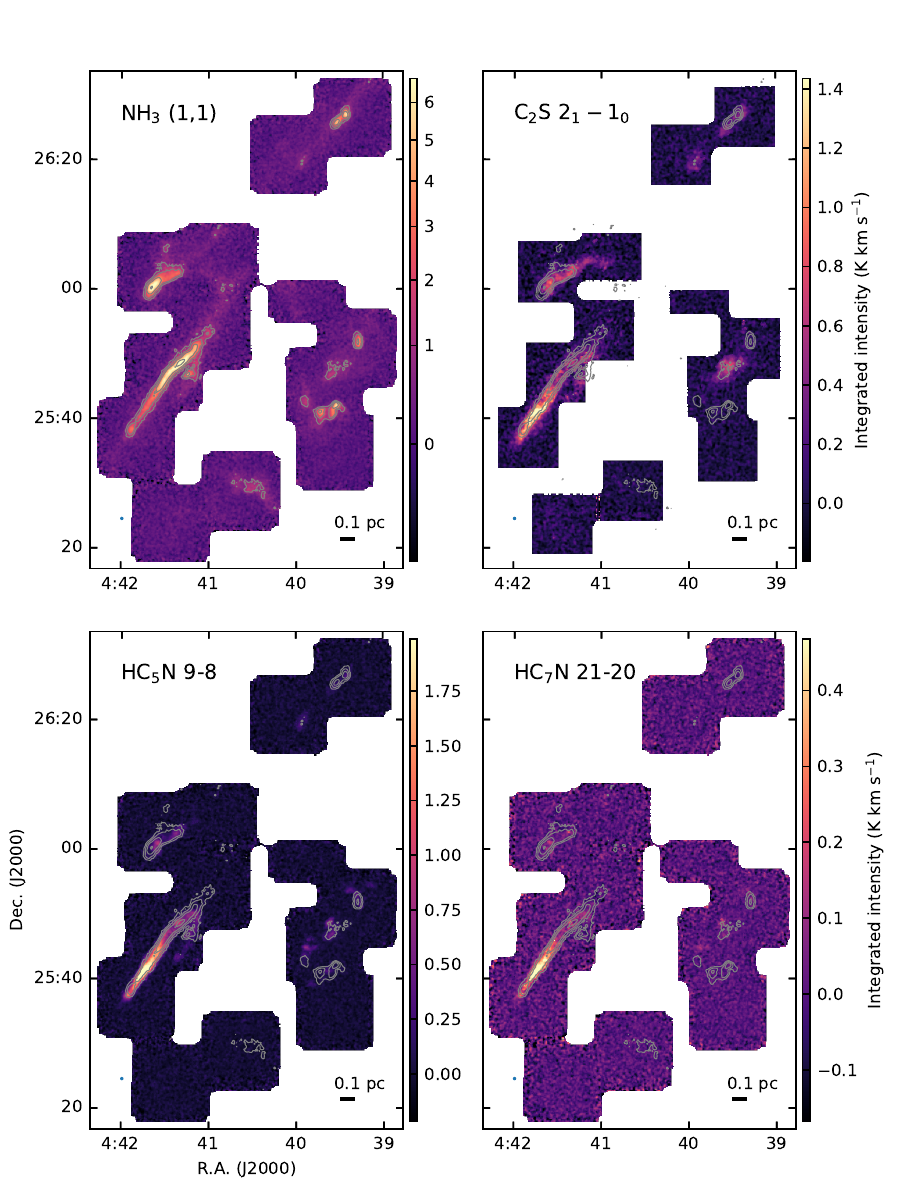}
    \caption{Integrated intensity of \amm\ (1,1), \ccs, \hcfn, and \hcsn\ emission toward Heiles Cloud 2 in Taurus. In all plots, grey contours follow the \amm\ (1,1) emission. The blue circle at lower left shows the GBT beam  \label{fig:hc2}}
\end{figure*}

Following the method described in more detail in \DRone, we use the \vlsr and \sigv results of the line fits 
to identify the windows where we compute the noise properties of the cubes. 
We show in Figure \ref{fig:hc2} the resulting integrated intensity maps of \amm(1,1), \ccs, \hcfn, and \hcsn toward Heiles Cloud 2. 

We detect \amm (3,3) above the S/N threshold toward six regions: L1688, NGC 1333, Orion A, NGC 2023-2024 and NGC 2067-2071 in Orion B, and Serpens Aquila. The emission is clearly associated with regions impacted by feedback from young intermediate or high-mass stars. In L1688, the emission is extended along and to the west of the Oph A clump, sometimes called $\rho$ Oph West \citep{Abergel1996-ISOCAM_Oph}, illuminated by the early B star HD 147889. In NGC 1333, \amm (3,3) is compact and located near the SVS 13 region, a hierarchical system of four protostellar sources \citep{Chen2009-PdBI_SVS13}, at least one of which is driving a large-scale chain of Herbig-Haro objects \citep{Chini1997-HH_Objects_YSOs}. \amm (3,3) is strongly detected toward Orion KL and Orion~A-S, with extended emission along the Orion Bar and north along the integral-shaped filament. Within NGC 2024, we find \amm (3,3) near the young high mass stars IRS2 and IRS2b. Toward Serpens Aquila, \amm (3,3) highlights multiple filamentary features seen in submillimeter continuum observations near the center of the W40 \ion{H}{2}  region \citep{Westerhout1958, Mallick2013-W40}. 

\subsection{Column densities of \ccs, \hcfn, and \hcsn\label{Sec:Ncol}}

For \amm, column density calculations are presented in \DRone\ based on modeling of the (1,1) and (2,2) inversion transitions. In this release, we present the \amm\ column density maps of all regions in the GAS survey, including those presented in DR1. 
Here, we determine the column densities for the additional species \ccs, \hcfn, and \hcsn\ where detected in the GAS survey. Table~\ref{tab:params} lists the rotational transitions and parameters for each line observed. 

We use the resulting fit line brightness and velocity dispersion in the following column density derivation. 
All three species are linear molecules. We therefore follow \citet{Mangum2015-Column_Density} to derive the total molecular column density, $N_\mathrm{tot}$, for rotational transitions $J_u \rightarrow J_u-1$ at frequency $\nu$ for optically thin lines, where
\begin{eqnarray}
    N_\mathrm{tot} = \frac{3h}{8\pi^3 S_{i,j} |\mu|^2} \frac{Q_\mathrm{rot}}{g_u}  \frac{\mathrm{exp}(\frac{E_u}{kT_{ex}})}{\mathrm{exp} {\left(\frac{h\nu}{kT_{ex}} \right)-1}} \int \tau_\nu \mathrm{d}v~.
\end{eqnarray}
Here, $\tau_\nu$ is the opacity as a function of frequency $\nu$. For these species, the associated line strength $S_{i,j} = J_u^2/(2J_u + 1)$, the rotational degeneracy $g_u = 2J_u + 1$, and $E_u = h B_0 J_u(J_u + 1)$ is the energy above ground state. 
For \hcfn and \hcsn, we calculate the rotational partition function, $Q_\mathrm{rot}$, following the approximation \citep{mcdowell_1988}
\begin{equation}
Q_\mathrm{rot} \simeq \frac{kT}{hB_0} \mathrm{exp}(hB_0/3kT) \,,
\end{equation}
which is accurate to 0.01\%  when $h B_0 / k T \lesssim 0.2$. For \ccs, we evaluate $Q_{rot} = \sum g_u \mathrm{exp}(- E_u/kT_{ex})$ over all $E_u/k < 1\,000$~K. For each species, the dipole moment $\mu$ and rotational constant $B_0$ were taken from the Spectral Line Atlas of Interstellar Molecules\footnote{\url{https://www.splatalogue.online}.} \citep[][for \hcfn, \hcsn, and  \ccs respectively]{alexander_1976,kroto_1978,saito_1987}. 

The integral over the opacity per frequency, ${\int \tau_\nu\, \mathrm{d}v}$, can be simplified to ${\sqrt{2\pi}}\sigma_v \tau$, where $\sigma_v$ is the {velocity dispersion}  of the emission line {assuming a Gaussian profile,} and
\begin{equation}
    \tau = -\ln \left[1 - \frac{T_B}{J(T_{ex}) - J(T_{bg})}\right]~, 
\end{equation}
is the {peak} opacity of the emission line. 
Here, $J(T)\equiv h\nu/k_B\, (\exp(h\nu/k_B\,T) - 1)$ is the Rayleigh-Jeans Equivalent Temperature, $T_B$ is the peak line brightness temperature, $T_{ex}$ is the excitation temperature, and $T_{bg} = 2.73$~K is the background temperature. 

We set $\tex  = 7$~K for all regions and all carbon-bearing species, with the exception of Heiles Cloud 2 in Taurus, where \citet{Smith2023-TMC1_Inflow_Fragments} determine $T_{ex} = 8.4$~K is more appropriate for \hcfn\ in TMC-1 using GAS data.  
Given the range of line brightness temperatures $T_B$ and velocity dispersions $\sigma_v$ where the carbon-bearing species are detected, we find that varying the assumed $T_{ex}$ between 5~K and 10~K produces an overall variation in the resulting column densities of $\lesssim 0.2$ dex in all species. As we do not have strong constraints on $T_{ex}$, we conclude this is our typical uncertainty in the reported values. 

\subsection{Molecular abundances and region masses}
\label{sec:cont}

The clouds observed with GAS were also mapped by the Herschel Gould Belt Survey \citep[HGBS;][]{Andre+10} 
in continuum emission from dust with both PACS \citep[70~\micron\ and 160~\micron;][]{Poglitsch2010-PACS}, 
and SPIRE \citep[250~\micron, 350~\micron, and 500~\micron;][]{Griffin2010-SPIRE}. 
\citet{Singh2022-Tau_Herschel} produced standardized modified blackbody fits of the spectral energy distribution (SED) across the HGBS clouds, producing maps of the dust opacity $\tau$ and temperature $T_d$ at 36\arcsec\ resolution (the angular resolution of Herschel in the 500~\micron\ passband)\footnote{\url{https://www.cita.utoronto.ca/HOTT/}}. 
We calculate the column density of \ce{H2}, \nh, following 
\begin{equation}
\tau_\nu = \kappa_\nu R_d \mu_{\ce{H2}} m_\mathrm{H}  \nh
\end{equation}
\citep[][Appendix D]{Singh2022-Tau_Herschel}, where $\kappa_\nu$ is the dust emission cross section per unit mass as a function of frequency $\nu$, $\tau_\nu$ is the dust opacity as a function of $\nu$, $R_d$ is the dust to gas mass ratio, $m_\mathrm{H}$ is mass of the hydrogen atom, and $\mu_{\ce{H2}} = 2.8$ \citep{Kauffmann2008-MAMBO}. 
To calculate \nh, we set $\kappa_\nu = 10$~cm$^2$~g$^{-1}$ at $\nu = 1$~THz, 
matching \cite{Singh2022-Tau_Herschel}, and $R_d = 0.01$. 
Because the angular resolution of the GAS and $N(\ce{H2})$ maps thus derived are very similar (32\arcsec\ and 36\arcsec, respectively), we assume that the column density maps are smooth at these angular scales and therefore regrid the resulting \nh  maps to match the GAS data. 
We are then able to determine per-pixel molecular abundances given the fit results of the various lines observed and \nh. 
We further calculate the total mass within the GAS mapped regions based on the \nh maps and the distances to each region, and show the results in Table \ref{tab:sources}. 

\subsection{\amm and continuum cross-matched core properties}

\citet{Pandhi2023-GAS_Shape_VGrad} used dendrogram analysis to identify dense molecular cores in integrated \amm\ (1,1) emission across most of the observed GAS clouds. They then cross-match the \amm-detected structures with core catalogs derived from submillimeter observations to produce a catalog of \amm\ cores that coincide with submillimeter cores, and we refer the reader to \cite{Pandhi2023-GAS_Shape_VGrad} for a detailed description of the core identification and matching analysis. Using the cross-matched catalog, \cite{Pandhi2023-GAS_Shape_VGrad} measure velocity gradients across the cores, examine the specific angular momentum as a function of core size, and show that the relative orientations of the cores, their specific angular momentum vectors, and the large-scale magnetic field as traced by Planck show little to no evidence for preferred alignment or anti-alignment. 

Here, we provide the mean and standard deviation of the gas properties derived from the \amm\ line fits measured within the \amm\ core boundaries for the cross-matched cores from \cite{Pandhi2023-GAS_Shape_VGrad}. We additionally calculate the virial parameter $\avir = M_\mathrm{vir}/M$ for each core, where $M$ is the continuum core mass. We define $M_\mathrm{vir}$ following \citet{Bertoldi_McKee_1992}:
\begin{equation}
M_\mathrm{vir} = \frac{5\sigma^2\,R}{a\,G}~,
    \label{eqn: Virial Mass}
\end{equation}
where $R$ is the core radius, $G$ is the gravitational constant, $\sigma$ is the total clump velocity dispersion, and $a$ is a factor that depends on the density profile of the core. 
For each core, we set $R$ equal to the continuum catalog core radius, and we set $a = 1$ \citep{Singh_2021_subvirial}{, 
which corresponds to a uniform density while for a free-falling core ($\rho(r) \propto r^{-1.5}$) $a = 1.25$}. 
The total clump velocity dispersion takes into account both the mean line width (given as \sigv) and the dispersion of the mean line-of-sight velocity, $\sigma_\mathrm{vlsr}$, across the core:
\begin{equation}
\sigma^2 = \sigma_\mathrm{vlsr}^2 + \sigma_v^2 - 
         \frac{k_\mathrm{B}\,\tkin}{m_{\amm}} + \frac{k_B\,\tkin}{\mu\, m_{\ce{H}}} ~,
    \label{eqn:sigma}
\end{equation}
where {we remove the \amm thermal line width and include the thermal line width of the mean gas,}  
\tkin\ is the mean kinetic temperature over the core, $k_B$ is the Boltzmann constant, $m_{\amm}$ is the molecular mass of ammonia, $m_{\ce{H}}$  
is the molecular mass of hydrogen, and $\mu = 2.37$ for molecular gas \citep{Kauffmann2008-MAMBO}. 
{This procedure calculates the non-thermal component of the core by removing the thermal velocity dispersion of the \amm line \citep[e.g.,][]{Pineda2021-B5_Ions_Neutral,Friesen2024-Serpens_South_EVLA}, 
The first term, $\sigma_\mathrm{vlsr}$, estimates the dense core bulk motion, which arises from variations of the
line-of-sight velocity corresponding to internal motions \citep[see][]{Singh_2021_subvirial}. 
Finally, it adds the thermal velocity dispersion for the average particle, which represents the thermal support within the dense core.}
Given $M_\mathrm{vir}$ for each core, we then calculate \avir. 

A short version of this core property catalog is shown in Table \ref{tab:core_cat}, while the full version is available online. 

\section{Discussion}
\label{sec:disc}

\subsection{\amm\ fit results across regions}

\begin{figure*}
    \includegraphics[width=0.98\textwidth]{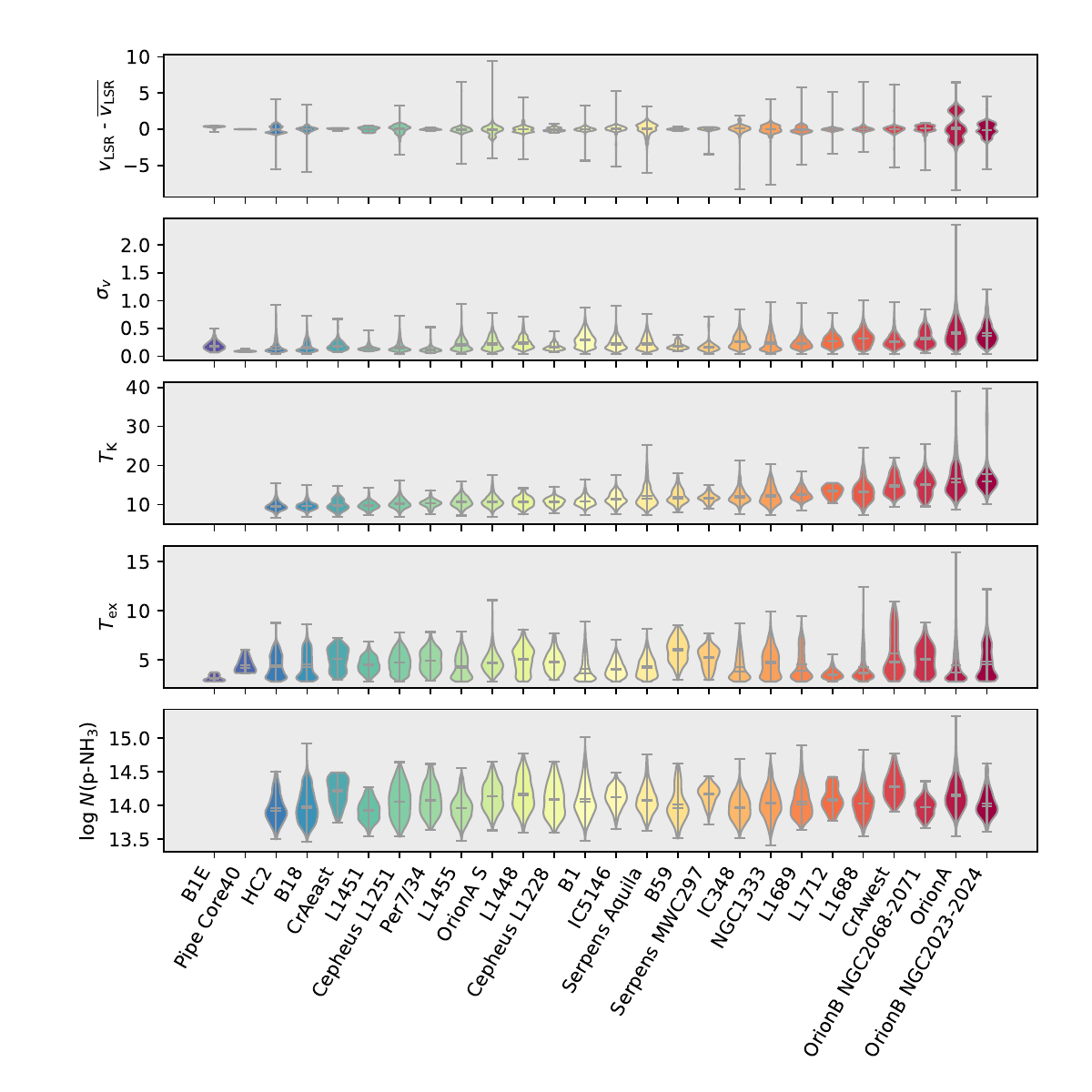}
    \caption{Fit parameter results for \amm\ (1,1) and (2,2) across all observed GAS regions. 
    Regions are ordered by increasing mean \tkin\ as measured by the hyperfine structure \amm\ fitting. 
    Violin plots show the extrema, median, and mean values (horizontal lines) and the distribution of values within each region. \label{fig:violins}}
\end{figure*}

Based on the \amm\ line fits, we identify several broad trends in the dense gas properties across the GAS clouds. For each region and fit parameter (\vlsr, \sigv, \tex, \tkin, and \namm), we list in Table \ref{tab:fitResults} the 50$^{th}$ percentile, as well as the 16$^{th}$ and 84$^{th}$ percentiles, of the distributions of fit parameters toward each observed region. We report percentiles rather than mean and 1-$\sigma$ values as the distributions are not necessarily Gaussian. We mask fit parameters based on the uncertainties in the \amm\ fits following \DRone. We additionally show in Fig.~\ref{fig:violins} the distribution of \amm\ fit parameters for the observed regions as violin plots. In these plots, the top and bottom bars show the maximum and minimum values for each parameter, while the mean and median values are shown as horizontal bars within the distribution. The fraction of pixels with a given parameter value is highlighted by the width of the violin. The regions are ordered from left to right and also colour-coded by increasing mean gas temperature \tkin, shown in the middle panel. For regions where fewer than 10 pixels have \tkin\ measurements after masking, we set the mean \tkin\ $= 10$~K. 

For most regions, \vlsr\ falls between $\sim 0$~\kms\ and $\sim 12$~\kms, with the exception of the two Cepheus clouds. While most regions do show a distribution of at least several \kms\ in \vlsr\ across the mapped areas, most of the gas often lies in a narrow \vlsr\ range (with the exception of Orion~A, and to a lesser extent NGC 2023-2024 in Orion~B). Regions with very little spread in \vlsr\ tend to be those with small mapped areas. 

Toward all regions, the observed velocity dispersion, \sigv, varies from very low, subsonic values to supersonic, but the range in \sigv\ values and the distribution change between regions. 
Fig.~\ref{fig:violins} shows that the variation in \sigv\ is clearly correlated with increased \tkin{, see also Appendix~\ref{sec:Tk_sigmav} for a direct comparison}. 
Colder regions tend to have a larger fraction of the dense gas with smaller \sigv, whereas warmer regions show an increasing fraction of the dense gas with larger line widths. We note that all regions still contain gas at very low \sigv, but these regions tend to be limited to compact cores in, e.g., Orion~A, rather than being broadly extended. 

Gas temperatures vary across the observed clouds, with median values ranging from $\sim 9$~K (e.g., HC2 in Taurus) to $\sim 16$~K (e.g., Orion~A and B). In general, regions that are colder on average also show a smaller spread in \tkin\ across the mapped area. 
While all regions still contain some fraction of the mapped area with very low \sigv, even when the mean value is higher, warmer regions do not necessarily retain similarly cold gas as those regions that are colder on average. Minimum gas temperatures, even in compact cores with low \sigv\ values, are higher by several K in regions like Orion~A compared with regions in Taurus. 

In contrast to \sigv\ and \tkin, both \tex\ and \namm\ show similar variation and spread across most of the regions. Mean \tex\ values tend to lie within $\sim 4 - 5$~K, with spreads of a few K. 

\floattable 
\begin{deluxetable}{lccccc|cc} 
\tabletypesize{\footnotesize} 
\tablecolumns{8} 
\tablewidth{0pt} 
\tablecaption{NH$_3$ fit results per region \label{tab:fitResults}} 
\tablehead{ 
\colhead{} & \colhead{$v_\mathrm{LSR}$} & \colhead{$\sigma_v$} & 
\colhead{$T_{ex}$} & \colhead{$T_K$} & \colhead{log $N$(NH$_3$)} & 
\colhead{$\sigma_{NT}$} & \colhead{log $X$(NH$_3$)} \\ 
\colhead{Region} & \colhead{(km~s$^{-1}$)} & \colhead{(km~s$^{-1}$)} & 
\colhead{(K)} & \colhead{(K)} & \colhead{(cm$^{-2}$)} & 
\colhead{(km~s$^{-1})$} & \colhead{} 
} 
\startdata 
B1 & 6.60$^{+0.25}_{-0.20}$ & 0.32$^{+0.21}_{-0.15}$ & 3.7$^{+1.8}_{-0.7}$ & 10.7$^{+1.3}_{-1.0}$ & 14.1$^{+0.3}_{-0.2}$ & 0.18$^{+0.11}_{-0.06}$ & -8.0$^{+0.2}_{-0.2}$\\ 
B18 & 6.19$^{+0.23}_{-0.29}$ & 0.15$^{+0.16}_{-0.06}$ & 4.3$^{+1.8}_{-1.2}$ & 9.5$^{+1.0}_{-0.8}$ & 14.0$^{+0.3}_{-0.2}$ & 0.09$^{+0.04}_{-0.02}$ & -8.1$^{+0.2}_{-0.2}$\\ 
B1E & 7.51$^{+0.12}_{-0.68}$ & 0.24$^{+0.90}_{-0.11}$ & 3.1$^{+0.5}_{-0.2}$ & \nodata & \nodata & \nodata & \nodata\\ 
B59 & 3.42$^{+0.13}_{-0.14}$ & 0.17$^{+0.09}_{-0.03}$ & 6.3$^{+1.6}_{-1.5}$ & 11.5$^{+2.6}_{-1.6}$ & 13.9$^{+0.3}_{-0.3}$ & 0.15$^{+0.10}_{-0.04}$ & -8.2$^{+0.2}_{-0.2}$\\ 
Cepheus\_L1228 & -8.11$^{+0.35}_{-0.17}$ & 0.16$^{+0.08}_{-0.03}$ & 4.7$^{+1.1}_{-1.0}$ & 10.7$^{+1.0}_{-1.2}$ & 14.1$^{+0.3}_{-0.2}$ & 0.14$^{+0.09}_{-0.03}$ & -8.0$^{+0.2}_{-0.2}$\\ 
Cepheus\_L1251 & -3.99$^{+0.36}_{-0.67}$ & 0.14$^{+0.14}_{-0.04}$ & 4.7$^{+1.2}_{-1.2}$ & 10.2$^{+1.4}_{-1.0}$ & 14.1$^{+0.3}_{-0.3}$ & 0.12$^{+0.11}_{-0.03}$ & -8.0$^{+0.2}_{-0.2}$\\ 
CrAeast & 5.66$^{+0.09}_{-0.14}$ & 0.18$^{+0.10}_{-0.06}$ & 5.1$^{+1.2}_{-1.4}$ & 9.6$^{+1.8}_{-1.0}$ & 14.2$^{+0.2}_{-0.2}$ & 0.14$^{+0.03}_{-0.05}$ & -8.0$^{+0.2}_{-0.2}$\\ 
CrAwest & 5.61$^{+0.25}_{-0.34}$ & 0.27$^{+0.16}_{-0.10}$ & 5.0$^{+3.4}_{-1.4}$ & 14.5$^{+3.4}_{-2.1}$ & 14.3$^{+0.2}_{-0.2}$ & 0.23$^{+0.12}_{-0.07}$ & -8.1$^{+0.2}_{-0.2}$\\ 
HC2 & 5.69$^{+0.57}_{-0.50}$ & 0.14$^{+0.20}_{-0.05}$ & 4.3$^{+1.5}_{-1.1}$ & 9.4$^{+1.3}_{-0.9}$ & 13.9$^{+0.3}_{-0.2}$ & 0.08$^{+0.05}_{-0.02}$ & -8.1$^{+0.2}_{-0.2}$\\ 
IC348 & 8.76$^{+0.38}_{-0.42}$ & 0.26$^{+0.17}_{-0.11}$ & 3.9$^{+1.9}_{-0.8}$ & 11.8$^{+2.3}_{-1.2}$ & 14.0$^{+0.2}_{-0.2}$ & 0.16$^{+0.11}_{-0.05}$ & -8.0$^{+0.2}_{-0.2}$\\ 
IC5146 & 3.87$^{+0.30}_{-0.27}$ & 0.22$^{+0.16}_{-0.09}$ & 4.0$^{+1.0}_{-0.7}$ & 11.1$^{+1.9}_{-1.5}$ & 14.1$^{+0.2}_{-0.2}$ & 0.19$^{+0.11}_{-0.07}$ & -7.8$^{+0.1}_{-0.2}$\\ 
L1448 & 4.44$^{+0.30}_{-0.39}$ & 0.24$^{+0.16}_{-0.09}$ & 5.1$^{+1.3}_{-1.6}$ & 10.7$^{+1.4}_{-1.3}$ & 14.2$^{+0.3}_{-0.2}$ & 0.18$^{+0.14}_{-0.05}$ & -8.0$^{+0.1}_{-0.2}$\\ 
L1451 & 4.37$^{+0.25}_{-0.37}$ & 0.14$^{+0.05}_{-0.02}$ & 4.5$^{+0.9}_{-0.7}$ & 9.6$^{+1.1}_{-1.0}$ & 13.9$^{+0.2}_{-0.2}$ & 0.11$^{+0.03}_{-0.02}$ & -7.8$^{+0.2}_{-0.2}$\\ 
L1455 & 5.14$^{+0.29}_{-0.28}$ & 0.21$^{+0.17}_{-0.09}$ & 4.3$^{+1.5}_{-1.1}$ & 10.6$^{+1.5}_{-1.2}$ & 14.0$^{+0.2}_{-0.2}$ & 0.12$^{+0.09}_{-0.04}$ & -7.9$^{+0.2}_{-0.2}$\\ 
L1688 & 3.47$^{+0.28}_{-0.22}$ & 0.33$^{+0.16}_{-0.14}$ & 3.8$^{+1.9}_{-0.6}$ & 13.1$^{+2.7}_{-2.4}$ & 14.0$^{+0.2}_{-0.2}$ & 0.17$^{+0.19}_{-0.08}$ & -8.2$^{+0.2}_{-0.2}$\\ 
L1689 & 4.03$^{+0.53}_{-0.33}$ & 0.24$^{+0.15}_{-0.08}$ & 4.0$^{+2.4}_{-0.9}$ & 12.4$^{+1.8}_{-1.1}$ & 14.0$^{+0.3}_{-0.2}$ & 0.17$^{+0.06}_{-0.05}$ & -8.2$^{+0.1}_{-0.1}$\\ 
L1712 & 4.74$^{+0.15}_{-0.12}$ & 0.28$^{+0.14}_{-0.10}$ & 3.5$^{+0.6}_{-0.4}$ & 13.0$^{+2.2}_{-2.0}$ & 14.1$^{+0.3}_{-0.1}$ & 0.28$^{+0.10}_{-0.07}$ & \nodata\\ 
NGC1333 & 7.72$^{+0.53}_{-0.49}$ & 0.25$^{+0.25}_{-0.13}$ & 4.7$^{+1.5}_{-1.4}$ & 11.9$^{+2.9}_{-1.8}$ & 14.0$^{+0.2}_{-0.2}$ & 0.17$^{+0.17}_{-0.08}$ & -8.0$^{+0.2}_{-0.2}$\\ 
OrionA & 8.40$^{+1.68}_{-1.26}$ & 0.52$^{+0.48}_{-0.27}$ & 3.8$^{+2.4}_{-0.8}$ & 15.6$^{+4.8}_{-2.9}$ & 14.1$^{+0.3}_{-0.2}$ & 0.30$^{+0.26}_{-0.13}$ & -8.1$^{+0.2}_{-0.2}$\\ 
OrionA\_S & 4.51$^{+0.46}_{-0.62}$ & 0.22$^{+0.16}_{-0.09}$ & 4.6$^{+1.1}_{-0.8}$ & 10.6$^{+1.6}_{-1.2}$ & 14.1$^{+0.2}_{-0.2}$ & 0.15$^{+0.09}_{-0.06}$ & -8.1$^{+0.2}_{-0.2}$\\ 
OrionB\_NGC2023-2024 & 10.15$^{+0.78}_{-0.87}$ & 0.41$^{+0.25}_{-0.17}$ & 4.6$^{+2.2}_{-1.5}$ & 16.0$^{+3.5}_{-1.5}$ & 14.0$^{+0.2}_{-0.1}$ & 0.26$^{+0.17}_{-0.09}$ & -8.1$^{+0.1}_{-0.1}$\\ 
OrionB\_NGC2068-2071 & 10.46$^{+0.49}_{-0.43}$ & 0.31$^{+0.21}_{-0.13}$ & 5.1$^{+1.5}_{-1.5}$ & 15.1$^{+2.5}_{-3.0}$ & 14.0$^{+0.1}_{-0.1}$ & 0.21$^{+0.18}_{-0.08}$ & -7.9$^{+0.1}_{-0.2}$\\ 
Perseus & 5.94$^{+0.11}_{-0.09}$ & 0.12$^{+0.04}_{-0.02}$ & 4.9$^{+1.2}_{-1.2}$ & 10.3$^{+0.9}_{-1.2}$ & 14.1$^{+0.2}_{-0.2}$ & 0.09$^{+0.04}_{-0.03}$ & -7.7$^{+0.2}_{-0.2}$\\ 
Pipe\_Core40 & 3.35$^{+0.03}_{-0.02}$ & 0.09$^{+0.02}_{-0.01}$ & 4.2$^{+1.3}_{-0.5}$ & \nodata & \nodata & \nodata & \nodata\\ 
Serpens\_Aquila & 7.29$^{+0.55}_{-0.74}$ & 0.23$^{+0.17}_{-0.10}$ & 4.2$^{+1.2}_{-0.9}$ & 11.3$^{+3.3}_{-1.7}$ & 14.1$^{+0.2}_{-0.2}$ & 0.15$^{+0.10}_{-0.05}$ & -7.9$^{+0.2}_{-0.2}$\\ 
Serpens\_MWC297 & 7.09$^{+0.10}_{-0.15}$ & 0.16$^{+0.08}_{-0.04}$ & 5.3$^{+0.9}_{-1.3}$ & 11.6$^{+1.2}_{-1.3}$ & 14.2$^{+0.1}_{-0.2}$ & 0.14$^{+0.06}_{-0.04}$ & -8.1$^{+0.1}_{-0.1}$\\ 
\enddata 
\tablecomments{For each parameter the values given are the 50$^{th}$ (16$^{th}$, 84$^{th}$) percentiles.
For B1E and Pipe\_Core40, there are not enough pixels to calculate the percentiles, therefore, we provide only the mean value.} 
\end{deluxetable} 

We further include in Table \ref{tab:fitResults} the median and (16$^{th}$, 84$^{th}$) values for the non-thermal velocity dispersion \signt, 
and the \amm\ abundance relative to \ce{H2}, \xamm. At each pixel where both \sigv\ and \tkin\ are fit well, we calculate \signt\ following
\begin{equation}
    \signt = \left(\sigma_v^2 - \frac{k_B T_K}{m_{\amm}}\right)^{1/2}~,
\end{equation}
where $m_\mathrm{\amm}$ is the molecular weight of \amm. 

We determine $\xamm = \namm/\nh$ at each pixel where \namm\ is fit well, using the \nh maps described in Section \ref{sec:cont}. We find that the individual region's average abundance, $\log_{10} X(\amm)$, is between $-$7.7 and $-$8.2, and a typical value of $-$8.0 is seen across all the clouds.
This value is consistent with previous observations of dense cores \citep{Tafalla+04,Friesen2009-NH3_Oph_highres,Pineda2022-HMM1_NH3_Depletion}, however, 
these observations do not have sufficient angular resolution to resolve possible \amm depletion as seen in other objects \citep{Caselli2022-L1544_NH2D_Depletion, Pineda2022-HMM1_NH3_Depletion,Lin2023-NH3_Depletion_Cores}.

\subsection{Appearance of subsonic structures}

\begin{figure*}
    \centering
    \includegraphics[width=\textwidth]{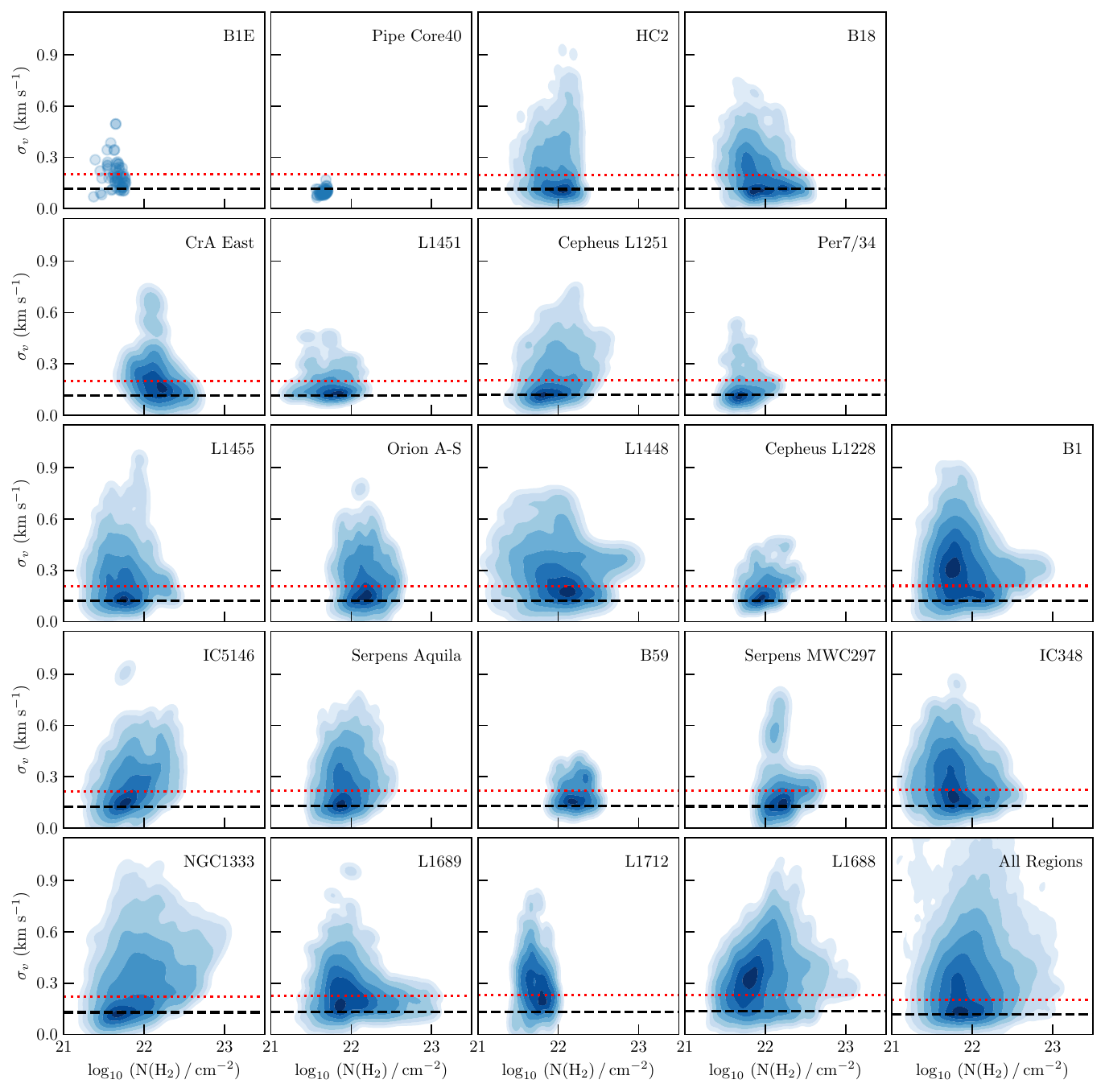}
    \caption{KDE of velocity dispersion as a function of \ce{H2} column density for all regions covered.
    The red-dotted and black-dashed lines correspond to the expected velocity dispersion, \sigv, 
    in the case of $\mathcal{M}_s$ equals 1 and 0.5, respectively, for the 
    median \tkin value of each region.
    In the case of B1E, Pipe Core40, and ``All Regions,'' we assume a temperature of 10 K. 
    Notice that since the regions are already sorted by the typical kinetic temperature, the horizontal  lines 
    have only a small variation between neighbouring panels.
    }
    \label{fig:dv_H2_KDE}
\end{figure*}

In the past, several works showed the presence of a transition to coherence, where 
the non-thermal velocity dispersion transitions from supersonic values in the molecular cloud down 
to subsonic values in the dense core \citep{Goodman1998-Transition_to_Coherence,Pineda2010-B5_Transition_to_Coherence,Hacar2018-ALMA_Orion,HChen-2019_Droplet,Choudhury2021-L1688_Extended}. 
We compare the velocity dispersion as a function of \ce{H2} column density for all regions in 
Fig.~\ref{fig:dv_H2_KDE}.
We use the kernel density estimation (KDE) implementation in \texttt{scipy} \citep{2020SciPy-NMeth},
while in regions with fewer than 100 detections, we plot the individual data points. 
The regions shown in Fig.~\ref{fig:dv_H2_KDE} are sorted by increasing mean 
kinetic temperature, \tkin, 
while the expected velocity dispersion for $\mathcal{M}_s=\signt/\csound$ equal 1 and 0.5 are 
marked by the red-dotted and black-dashed horizontal lines, respectively.
This figure shows that there is no {\it universal column density} at which the 
level of non-thermal velocity dispersion is subsonic. 

{Furthermore, we compute the KDE of the effective radius and the minimum column density for all subsonic structures. 
Independent structures are identified in the images using \texttt{scipy.ndimage.label}, and only those with at least 10 pixels are retained. 
For each structure, we determine the effective radius and the minimum \ce{H2} column density, and then calculate the KDE across all regions. 
The radius distribution (Fig.~\ref{fig:subsonic_KDE}-left) shows that subsonic cores are typically compact ($\approx0.05$~pc), although some extend up to 0.3~pc. 
Similarly, the column density distribution (Fig.~\ref{fig:subsonic_KDE}, right) exhibits a peak at $\approx 6\times10^{21}$~\sqcm with a FWHM of $\approx 0.5$~dex, while spanning more than one order of magnitude (1 dex). 
Although both properties display characteristic values, neither can be described by a universal value. 

\begin{figure*}[ht!]
    \centering
    \includegraphics[width=0.45\textwidth]{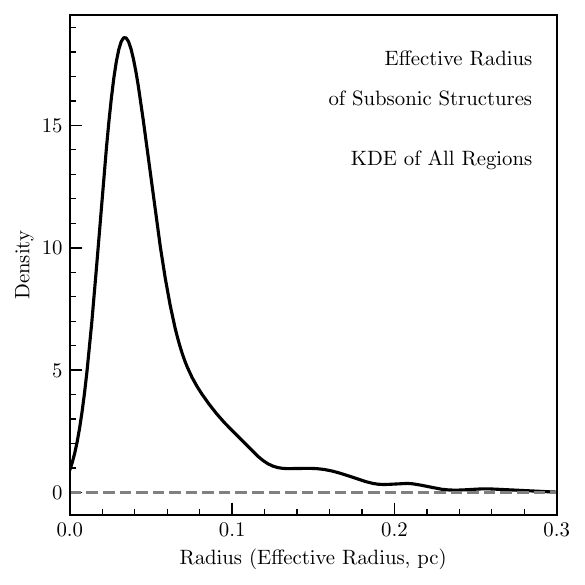}
    \includegraphics[width=0.45\textwidth]{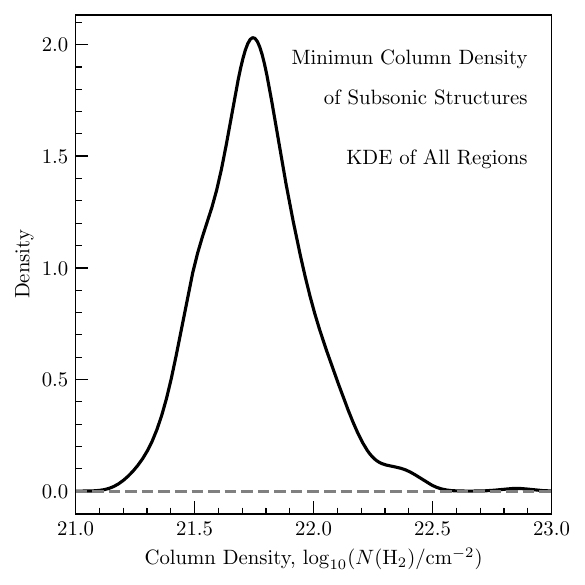}
    \caption{KDE of the effective radius and minimum column density of subsonic structures across all regions. 
    In both cases, the distributions exhibit peaks (representative values) but span a wide range, indicating the absence of a universal value for these structures. 
    {\it Left:} Distribution of the effective radius ($R_{\mathrm{eff}} = \sqrt{A/\pi}$) for all subsonic structures. 
    It shows a peak at $\approx 0.05$~pc, with a pronounced tail extending up to 0.3~pc. 
    {\it Right:} Distribution of the minimum \ce{H2} column density within subsonic structures. 
    It shows a peak near $6\times 10^{21}$\,\sqcm and covers a broad range, spanning more than an order of magnitude.}
    \label{fig:subsonic_KDE}
\end{figure*}
}

\begin{figure*}[hbt!]
    \centering
    \includegraphics[width=0.9\textwidth]{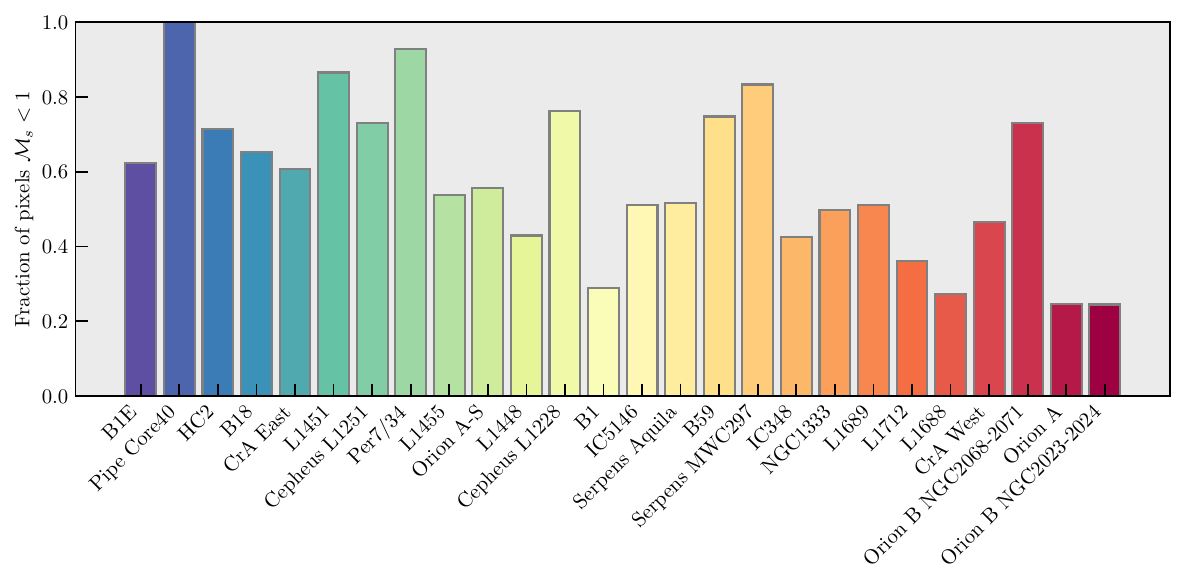}
    \caption{Fraction of subsonic non-thermal velocity dispersion present in the different regions. 
    In the case of B1E and Pipe Core40, we assume a temperature of 10 K.
    They are ordered and sorted by increasing mean \tkin, as in Fig.~\ref{fig:violins}.}
    \label{fig:fraction_subsonic}
\end{figure*}

We compare the fraction of pixels displaying a $\mathcal{M}_s<1$ in Fig.~\ref{fig:fraction_subsonic}.
This fraction is determined by using the mean temperature in the cloud for all pixels 
with a good velocity dispersion determination. 
The fraction is correlated with mean \tkin in the region, although this depends on the average density in the region 
and the level of star-formation in the region, and it shows that even in active star-forming regions 
quiescent {areas} (likely cores) are present.
{This suggests that stellar feedback plays a role in the scales covered with these observations, see also \cite{Friesen2024-Serpens_South_EVLA}.}

{Simulations further support this interpretation. 
The effect of feedback is studied by \cite{Neralwar2024-STARFORGE_Feedback_Cores}, where they find that cores subject to stronger feedback are generally more compact but exhibit systematically higher velocity dispersions and virial parameters, implying they are more gravitationally unbound. 
\cite{Offner2022-Core_Evolution} show that many dense structures nevertheless pass through a coherent, low-turbulence phase that is consistent with the quiescent cores identified in observations such as GAS, although not all such cores survive to form stars. 
Taken together, these results suggest that while feedback injects turbulence and unbinds a large fraction of dense gas, it simultaneously permits the emergence of coherent, quiescent cores within active star-forming regions, providing a natural explanation for the subsonic pockets seen in our data.}

\subsection{Regions with cyanopolyyne and \ccs\ detections}

We detect \hcfn\ above a S/N $=3$ toward 13 separate regions within the survey. 
Table~\ref{tab:hcfnResults} provides a list of the regions with significant \hcfn\ detection, and the 50$^{th}$ percentile, as well as the 16$^{th}$ and 84$^{th}$ percentiles of the column densities \nhcfn and \nh, and the \hcfn\ abundance relative to \ce{H2}, $X(\hcfn/\ce{H2})$, at the locations where \hcfn\ is detected. 
We find typical mean $\nhcfn \sim$ a few $\times 10^{12}$~\sqcm and mean $X(\hcfn/\ce{H2})$ values of a few $\times 10^{-10}$, and variations in the mean values between regions are within a factor of ten. 
In Table~\ref{tab:hcfnResults} we further list the 50$^{th}$ percentile, as well as the 16$^{th}$ and 84$^{th}$ percentiles of the column density of \amm\ and the abundance ratio $N(\hcfn)/N(\amm)$ in pixels where both lines are detected. 
The highest abundances of \hcfn\ relative to \amm\ are found in Heiles Cloud 2 in Taurus, 
which contains the well-studied carbon-rich TMC-1 region \citep[e.g.,][]{Suzuki1992-CCS_NH3,Kaifu2004-TMC1_Survey}. 
Toward the other GAS-targeted region in Taurus, B18, we also see extended \hcfn\ emission that is only slightly lower in relative abundance than toward HC2. 
Overall, where \hcfn\ emission is detected, the abundances found are in general agreement with previous studies 
of low-mass star-forming regions \citep[e.g.,][]{BensonMyers1983-HC5N,Suzuki1992-CCS_NH3,Codella1997-TMC_NH3_HC5N,Hirota2009-Carbon_Chain_Cores}. 
It is difficult to make direct comparisons with detailed studies of particular regions, such as TMC-1, as we observed only one \hcfn\ transition and thus have made simple but well-motivated assumptions about the line excitation when calculating column densities and abundances. 

Of the other carbon-bearing species, we detect \ccs\ toward a small subset of regions where \hcfn\ is detected: B1 in the Perseus molecular cloud, B18 and HC2 in the Taurus molecular cloud, and Serpens Aquila. We find no detections in regions where \hcfn\ is not detected. 
As noted in Section \ref{sec:data}, the \ccs\ maps have higher rms noise values, which likely limited the number of regions with detections. The cyanopolyyne \hcsn\ was only detected toward B1, HC2, and Orion~A in the $J = $(21--20) line, and toward HC2 only in the $J = $(22--21) line. We present similar abundance results as in Table \ref{tab:hcfnResults} for C$_2$S and HC$_7$N in Section \ref{sec:app_col}, where detected. 

\floattable 
\begin{deluxetable}{lccccc} 
\tabletypesize{\footnotesize} 
\tablecolumns{6} 
\tablewidth{0pt} 
\tablecaption{Regions with HC$_5$N detections \label{tab:hcfnResults}} 
\tablehead{ 
\colhead{Region} & \colhead{log $N$(HC$_5$N)} & 
\colhead{log $N$(NH$_3$)} & 
\colhead{log $X$(HC$_5$N/NH$_3$)} & 
\colhead{log $N$(H$_2$)} & 
\colhead{log $X$(HC$_5$N/H$_2$)} } 
\startdata 
B1 & 12.44$^{0.27}_{-0.20}$ & 14.25$^{0.35}_{-0.30}$ & -1.81$^{0.40}_{-0.33}$ & 21.05$^{0.47}_{-0.33}$ & -9.63$^{0.35}_{-0.39}$ \\ 
B18 & 12.65$^{0.40}_{-0.28}$ & 14.02$^{0.28}_{-0.22}$ & -1.33$^{0.43}_{-0.33}$ & 21.18$^{0.29}_{-0.21}$ & -9.38$^{0.36}_{-0.39}$ \\ 
Cepheus L1228 & 12.59$^{0.31}_{-0.23}$ & 14.01$^{0.40}_{-0.17}$ & -1.51$^{0.51}_{-0.37}$ & 20.99$^{0.30}_{-0.27}$ & -9.35$^{0.36}_{-0.26}$ \\ 
Cepheus L1251 & 12.53$^{0.26}_{-0.20}$ & 14.19$^{0.23}_{-0.27}$ & -1.63$^{0.33}_{-0.40}$ & 20.98$^{0.36}_{-0.17}$ & -9.54$^{0.23}_{-0.28}$ \\ 
CrAeast & 12.92$^{0.27}_{-0.22}$ & 14.34$^{0.12}_{-0.16}$ & -1.39$^{0.41}_{-0.30}$ & 21.39$^{0.36}_{-0.22}$ & -9.26$^{0.26}_{-0.39}$ \\ 
HC2 & 12.72$^{0.49}_{-0.32}$ & 13.94$^{0.29}_{-0.18}$ & -1.09$^{0.60}_{-0.36}$ & 21.49$^{0.21}_{-0.24}$ & -9.31$^{0.43}_{-0.31}$ \\ 
IC5146 & 12.68$^{0.02}_{-0.11}$ & 14.26$^{0.02}_{-0.04}$ & -1.60$^{0.03}_{-0.03}$ & 21.19$^{0.39}_{-0.28}$ & -9.59$^{0.01}_{-0.11}$ \\ 
L1448 & 12.63$^{0.18}_{-0.21}$ & 14.37$^{0.23}_{-0.19}$ & -1.72$^{0.18}_{-0.29}$ & 20.96$^{0.38}_{-0.23}$ & -9.61$^{0.22}_{-0.37}$ \\ 
L1451 & 12.50$^{0.35}_{-0.15}$ & 14.13$^{0.09}_{-0.19}$ & -1.30$^{0.20}_{-0.19}$ & 21.02$^{0.40}_{-0.21}$ & -9.26$^{0.27}_{-0.18}$ \\ 
NGC1333 & 12.34$^{0.26}_{-0.25}$ & 14.18$^{0.10}_{-0.19}$ & -1.85$^{0.31}_{-0.21}$ & 21.07$^{0.39}_{-0.22}$ & -9.68$^{0.21}_{-0.22}$ \\ 
OrionA & 12.60$^{0.23}_{-0.22}$ & 14.38$^{0.25}_{-0.30}$ & -1.73$^{0.32}_{-0.24}$ & 21.24$^{0.49}_{-0.41}$ & -9.69$^{0.33}_{-0.26}$ \\ 
OrionA S & 12.73$^{0.30}_{-0.21}$ & 14.22$^{0.20}_{-0.27}$ & -1.42$^{0.29}_{-0.29}$ & 21.34$^{0.49}_{-0.35}$ & -9.48$^{0.23}_{-0.26}$ \\ 
Serpens Aquila & 12.72$^{0.25}_{-0.19}$ & 14.23$^{0.32}_{-0.23}$ & -1.56$^{0.34}_{-0.22}$ & 21.56$^{0.21}_{-0.17}$ & -9.26$^{0.24}_{-0.20}$ \\ 
\enddata 
\tablecomments{Values given are the 50$^{th}$ percentile, along with the 16$^{th}$ and 84$^{th}$ percentiles of the parameter distributions for all columns. Column densities and abundances of \hcfn relative to H$_2$ are calculated over all pixels with good line fits, while the \amm column densities and \hcfn abundances relative to \amm are calculated in regions where both lines are detected.}
\end{deluxetable} 

\hcfn\ and \ccs\ emission is detected in both extended and compact (approximately beam-sized) features 
within the different regions. These differences in morphology may point to the molecules' origins in the cold, 
dense gas in some locations, as well as through warm carbon-chain chemistry near protostellar sources in others. 
To highlight the varying distributions of the molecular tracers in one region, Figure \ref{fig:hc2} shows 
the integrated intensity maps for \amm\ (1,1), \ce{CCS}, \ce{HC5N}, and \ce{HC7N}  toward Heiles Cloud 2, containing the TMC-1 filament. 
Here, we have overlaid \amm\ (1,1) integrated intensity contours on the emission maps of the 
carbon-bearing species to compare their distributions better. 
The offset between the cyanopolyyne and \amm\ peaks in TMC-1 is well-known \citep{Little1979-HC2_Carbon_NH3}, 
but the Figure shows that much of the extended carbon-chain emission elsewhere in the region frequently lies 
close to extended \amm\ emission, but offset from \amm\ emission peaks. 
In contrast, compact, bright \ce{HC5N} emission that is spatially coincident with compact \amm\ emission is seen toward the L1451 region, specifically toward the L1451-west source previously identified in \dia\ emission \citep{Storm2016-CLASSY_L1451}. 
\cite{Storm2016-CLASSY_L1451} argue that L1451-west is a highly evolved core that is potentially still starless. 
In this scenario, however, we expect much of the carbon-bearing species to be frozen out onto dust grains, which is difficult to reconcile with the bright \ce{HC5N} emission. Nevertheless, at the resolution of the GAS observations, the source has a mean \tkin\ $= 9.6$~K, making this an interesting and enigmatic core. We do not detect \ce{CCS} or \ce{HC7N} at this location.

In some regions, spatial and kinematic offsets between the carbon-chain species and \amm have been explained in terms of relative dynamical ages \citep[e.g., TMC-1;][]{Suzuki1992-CCS_NH3}, and as tracers of mass infall or accretion onto molecular filaments \citep{Friesen2013-Serpens_South_Infall,Smith2023-TMC1_Inflow_Fragments}. 
A detailed region-by-region analysis of these features is beyond the scope of this paper. 
In the next section, however, we focus on \ce{CCS} and \ce{HC5N} emission in the B1 region, 
and show how the carbon-chain molecular emission extends a previously-detected infalling streamer 
of gas from small (disk) scales to clump scales. 

\subsection{Origin of Streamers: Kinematical Connection}
Recently, interferometric observations of young stellar objects revealed
the presence of asymmetric infall along linear ``streamers.''
These structures deliver mass from beyond the dense core 
down to the disk 
\citep[or disk scales, see][]{Garufi2022-Streamer,Pineda2020-Streamer,Thieme2022-Lupus_Streamer,ValdiviaMena2022-Streamer,Hsieh2023-SVS13A_Prodige,ValdiviaMena2023-B5_HC3N_Streamer,Flores2023-eDisk_IRS63,Speedie2025-Streamer_ABAur}. 
Unfortunately, most observations detect these streamers {only} up to the angular extents 
of the interferometers' primary beam.

\begin{figure*}
    \centering
    \gridline{\fig{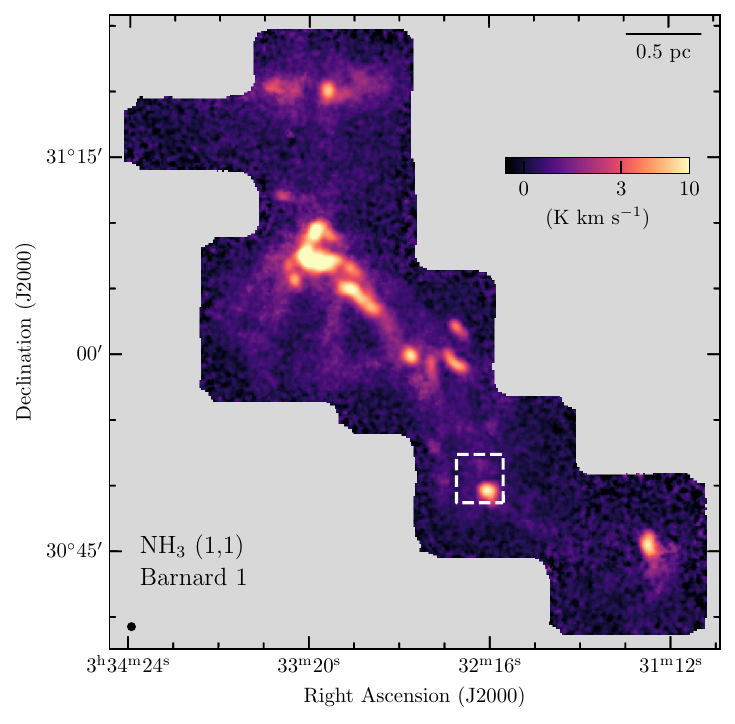}{0.62\textwidth}{(a)}}
\gridline{\fig{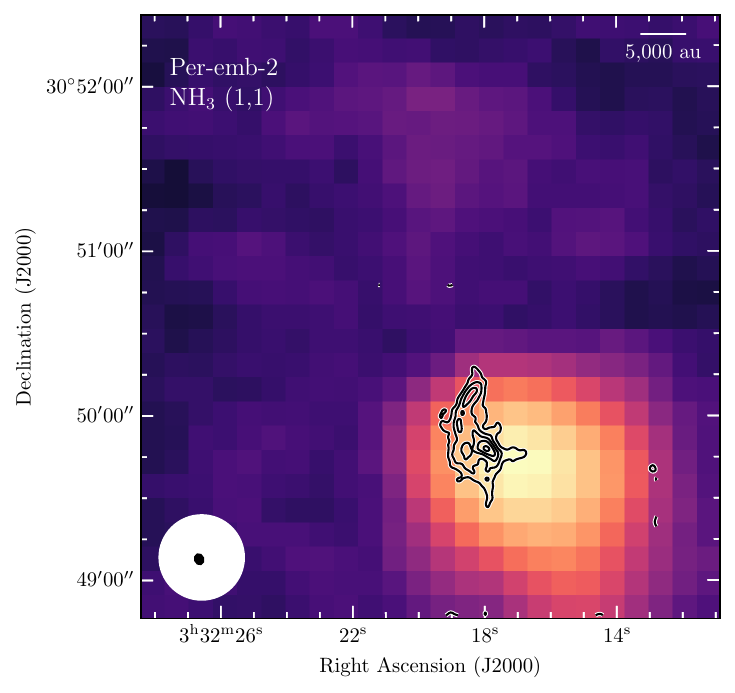}{0.43\textwidth}{(b)} 
          \fig{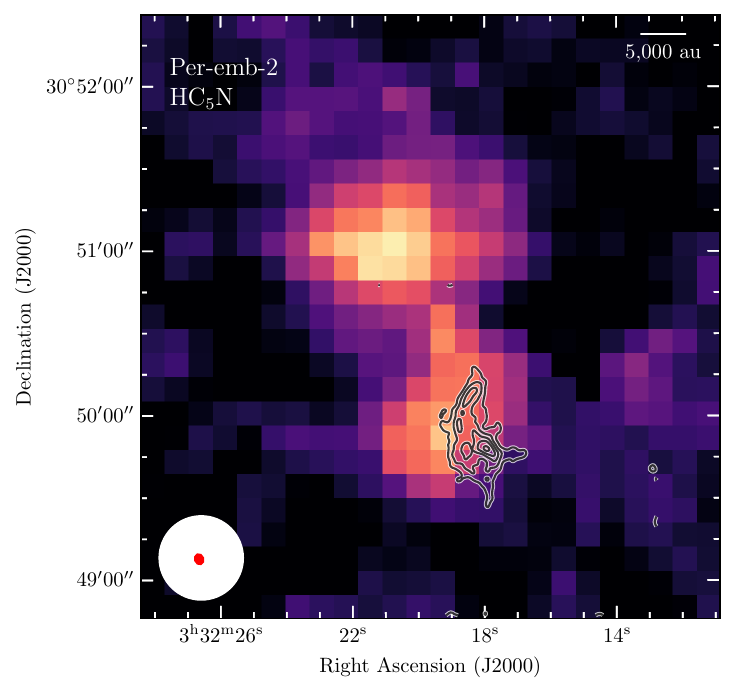}{0.43\textwidth}{(c)}}
    \caption{Extension of the streamer in  \hcfn beyond the \amm core.
    Panel a shows the large area \amm (1,1) map of the 
    Barnard 1 region, which includes Per-emb-2. 
    The dashed box marks the zoom-in region around Per-emb-2.
    Panels b and c show the zoom-in of the \amm (1,1) and 
    \ce{HC5N} emission, respectively, overlaid with the 
    \ce{HC3N} (10--9) integrated intensity from 
    NOEMA \citep{Pineda2020-Streamer}. 
    The beam sizes and scale bars are shown in the bottom left and top right 
    corners, respectively.
    }
    \label{fig:B1_TdV}
\end{figure*}

Our large-area maps reveal a new region in Perseus with emission from \ccs  and \hcfn that
extends beyond a previously observed streamer. 
We show in Fig.~\ref{fig:B1_TdV}-a the large area \amm(1,1) map 
of the Barnard 1 region, which includes the core hosting the YSO, Per-emb-2.
We zoom in into the Per-emb-2 YSO in the \amm and \hcfn emission 
in Fig.~\ref{fig:B1_TdV}-b and -c, respectively. 
These Figures show that \amm is related to the dense core, 
while the \hcfn traces a different structure, 
an extension to a previously seen streamer shown in the contours.
This spatial connection suggests that the streamer is linked to the larger gas reservoir seen in our data, 
see also \cite{Taniguchi2024-Per_emb_2_SLED} for a more detailed abundance determination.

\begin{figure*}
    \centering
    \includegraphics[width=0.4\textwidth]{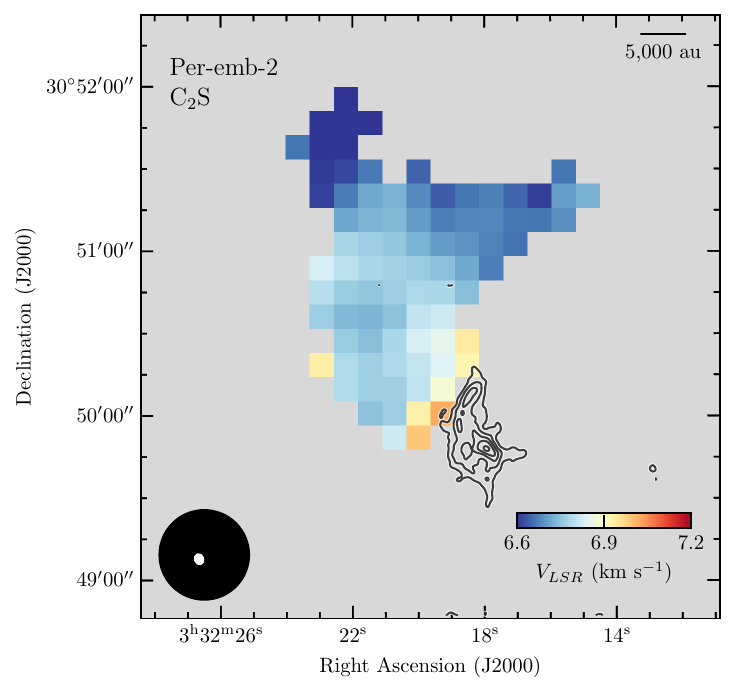}
    \includegraphics[width=0.4\textwidth]{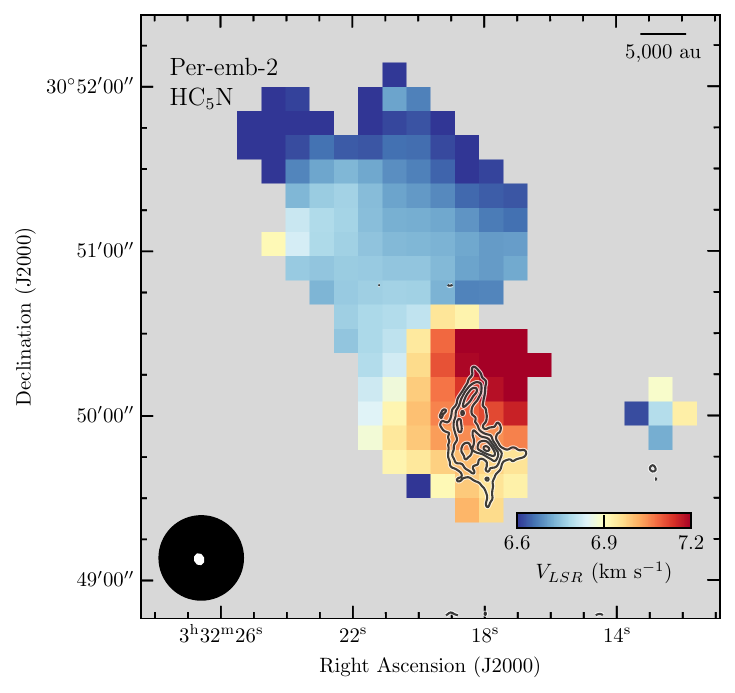}
    \caption{Velocity maps obtained towards Per-emb-2 from a single Gaussian fit to the \ccs  and 
    \hcfn data cubes from GAS. 
    A clear velocity gradient connects the streamer and the main gas reservoir.
    Overlaid in contours is the \ce{HC3N} (10--9) integrated intensity from 
    NOEMA \citep{Pineda2020-Streamer}.
    The beam sizes and scale bars are shown in the bottom left and top right 
    corners, respectively.}
    \label{fig:B1_Vc}
\end{figure*}

We derive the velocity maps for \ccs and \hcfn data from GAS with Gaussian fits and 
compare them to the extent of the NOEMA-identified streamer in Fig.~\ref{fig:B1_Vc}.
The emission from \ccs and \hcfn displays similar velocity maps, with 
clear blue velocities at large distances ($>$10,000 au) and a redder velocity 
component close to the Per-emb-2 source.
The velocity gradient is smooth, suggesting that this extended emission
detected in the GAS data traces the likely origin of the streamer.

\begin{figure*}
    \centering
    \includegraphics[width=0.45\textwidth]{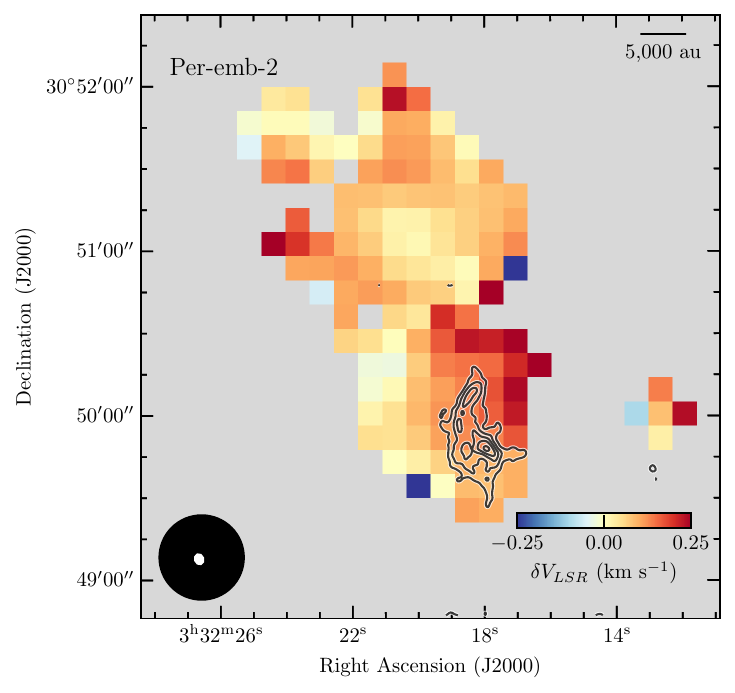}
    \includegraphics[width=0.45\textwidth]{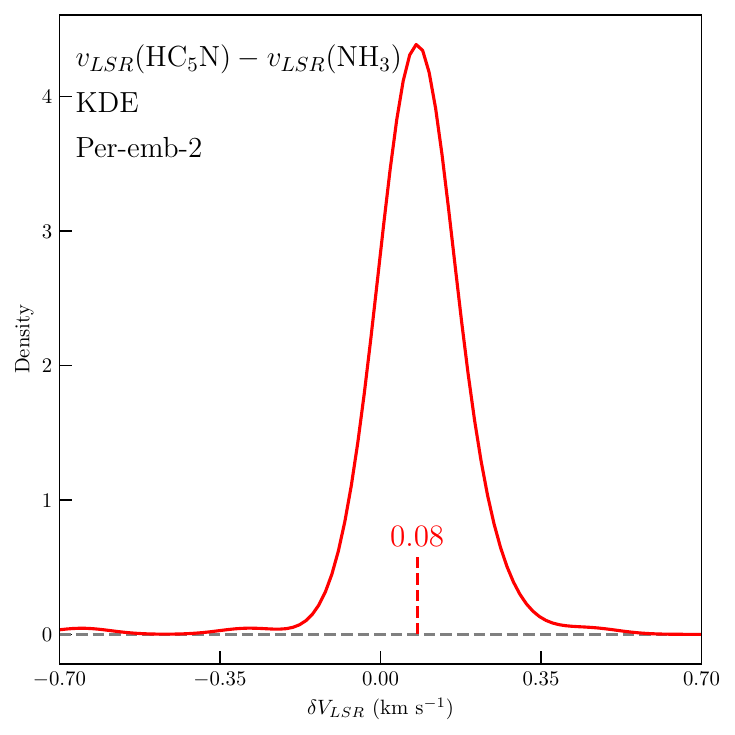}
    \caption{Velocity differences between \hcfn and \amm line emission.
    Left panel shows the velocity difference between a single Gaussian fit to the \hcfn line and the hyperfine fit of \amm obtained by GAS. 
    A clear velocity gradient connects the streamer and the main gas reservoir.
    Overlaid in contours is the \ce{HC3N} (10--9) integrated intensity from 
    NOEMA \citep{Pineda2020-Streamer}.
    The beam sizes and scale bars are shown in the bottom left and top right 
    corners, respectively.
    The right panel shows the KDE of the velocity difference in the regions shown in the left panel.
    The vertical line shows the median value of the velocity difference.
    }
    \label{fig:B1_delta_Vc}
\end{figure*}

We calculate the velocity difference between \hcfn and \amm, $\delta V_{\mathrm{LSR}}=\vlsr(\hcfn) - \vlsr(\amm)$, 
obtained from the respective line fits and where there is no evidence for multiple velocity components along the line of sight. 
The difference map and its KDE are shown in Fig.~\ref{fig:B1_delta_Vc}.  
The velocity difference map shows that the \hcfn is systematically red-shifted with respect to \amm by 
0.08~\kms. 
Similar velocity differences were also reported in Serpens-South \citep{Friesen2013-Serpens_South_Infall} 
between \hcsn and \amm. 
These velocity differences suggest that these molecules are not only tracing different volumes, 
but they also trace infalling motions of more chemically fresh material \citep{Pineda2020-Streamer,ValdiviaMena2024-ProPStar_Streamer_Survey}.

{Another possibility to explain the velocity difference is the presence of infalling or expanding motions, 
which are previously seen in different optically thick lines \citep{Anglada1987-Infall_Spectra,Mardones1997_Infall,Lee_1999_infall,Lee_2004_infall,Friesen2013-Serpens_South_Infall}.
We inspected the emission of \hcfn in the regions shown in Fig.~\ref{fig:B1_delta_Vc}, and we find that the line profiles 
are well fitted with a single Gaussian profile without evidence for an asymmetry in them.
We also compared the typical difference in the observed velocity dispersion difference (0.02 \kms)
and the non-thermal velocity dispersion (0.008 \kms), and find that they are comparable or smaller than the typical 
uncertainty in the observed velocity dispersion (0.02--0.03 \kms).
Therefore, these observations do not support the opacity effect in an expanding or contracting cloud to explain the observations.
}

\section{Summary}
\label{sec:summary}

Here, we present the final data release, DR2,  from the Green Bank Ammonia Survey, where we have mapped a large sample of northern hemisphere star-forming clouds in emission from \amm (1,1), (2,2), and (3,3), along with \hcfn, \hcsn, and \ccs. 
We have described the observations, calibration, and combination of the datasets, as well as, 
the line fitting of all observed species. 
The cubes, moment maps, and maps of the fitted parameters are publicly available. 

\begin{itemize}
    \item We use the \amm single velocity component fit to determine the physical properties of the dense gas in the regions covered. The kinetic temperature, \tkin, appears correlated with the mean value of the \sigv in the region, 
    as well as the maximum value of \sigv. 
    This trend points to the widespread role of feedback from recent the star-formation on larger scales.
    \item We combine the kinetic temperature and velocity dispersion determinations to estimate the 
    non-thermal velocity dispersion across the regions. 
    {We find that regions with subsonic levels of non-thermal velocity dispersion are commonly present 
    across different star forming regions.}
    We find that the fraction of pixels with 
    subsonic levels correlates well with the median kinetic temperature in the region. 
    Even in active regions, like Orion~A or Orion~B, approximately 20\% of the pixels display subsonic 
    levels of non-thermal velocity dispersion.
    \item We do not find a typical or universal column density at which the non-thermal velocity dispersion 
    is mostly subsonic.
    \item We find that the $\log_{10} X(\amm)$ in individual region varies between $-$7.7 and $-$8.2, with a typical value of $-$8.0 when taking all clouds into account. We do not detect evidence for \amm depletion, likely due to the angular resolution of the GAS observations.
    \item The typical abundance ratio between \hcfn and \amm ranges from -1.8 and -1.0 dex.
    \item We generate a core catalog of structures identified in \amm using a \texttt{dendrograms} analysis 
    and which are matched with continuum catalogs.
    \item We explore the possibility of finding the origin of streamers by focusing on Per-emb-2 and comparing 
    the relative \hcfn and \amm centroid velocities. We find that \hcfn is systematically red-shifted with respect to 
    \amm {in this source}. 
\end{itemize}

\begin{acknowledgments}
    The authors thank the Green Bank Observatory staff who supported this project. 
    The Green Bank Observatory is a facility of the National Science Foundation operated under cooperative agreement by Associated Universities, Inc.
    Part of this work was supported by the Max-Planck Society. 
    Y. Shirley was partially supported by an Astronomy and Astrophysics Grant AST-2205474 from the National Science Foundation.
    HK acknowledges support from an NSERC Discovery Grant.
    {AG acknowledges support from the NSF under grant CAREER 2142300.
    SO acknowledges support from a Peter O'Donnell Distinguished Researcher Fellowship and a Donald Harrington Fellowship.
    }
    This research used the Canadian Advanced Network For Astronomy Research \citep[CANFAR;][]{Gaudet2010-CANFAR}
     operated in partnership by the Canadian Astronomy Data Centre and The Digital Research Alliance of Canada 
     with support from the National Research Council of Canada the Canadian Space Agency, CANARIE and 
     the Canadian Foundation for Innovation.
\end{acknowledgments}

\facility{GBT}
\software{Astropy \citep{astropy:2013, astropy:2018}, 
pyspeckit \citep{pyspeckit2011, pyspeckit2022},
matplotlib \citep{Hunter_2007},
spicy \citep{2020SciPy-NMeth} }

\appendix

\section{Map noise levels\label{appendix:rms}}

\noprint{\figsetstart}
\noprint{\figsetnum{10}}
\noprint{\figsettitle{Histograms of rms noise.}}

\figsetgrpstart
\figsetgrpnum{10.1}
\figsetgrptitle{B1}
\figsetplot{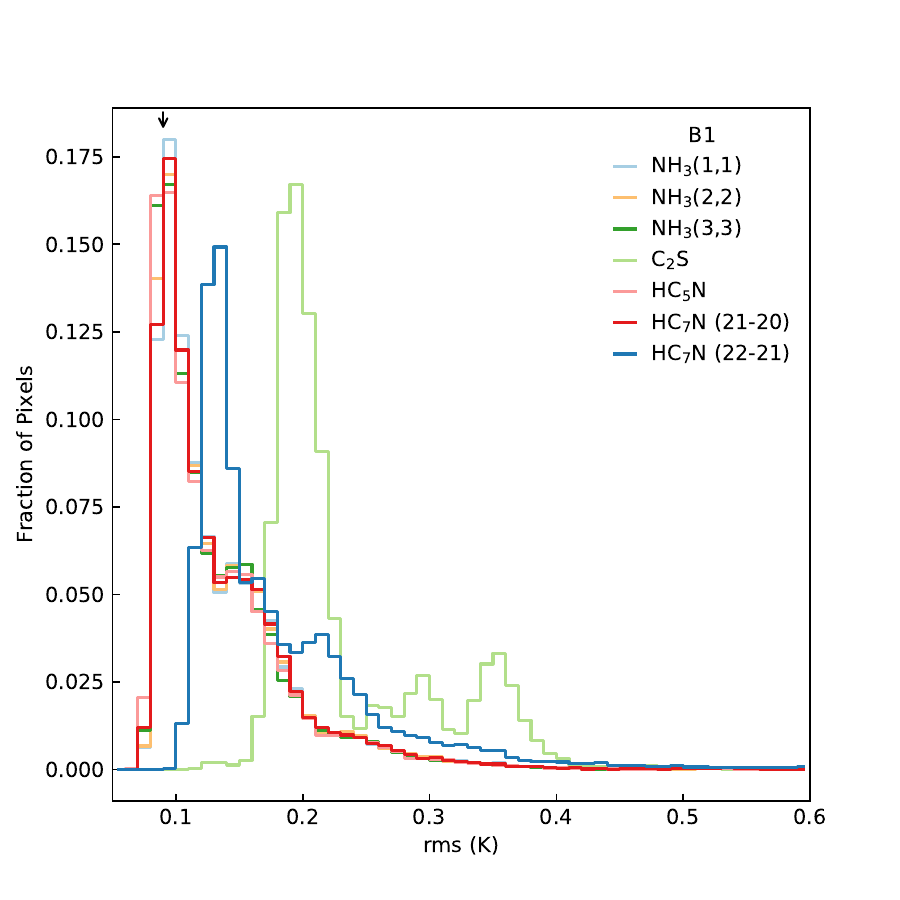}
\figsetgrpnote{B1 histogram of noise values per pixel for all observed lines. The black arrow shows the mode of the rms values for the NH3 (1,1) spectral window.}
\figsetgrpend

\figsetgrpstart
\figsetgrpnum{10.2}
\figsetgrptitle{B18}
\figsetplot{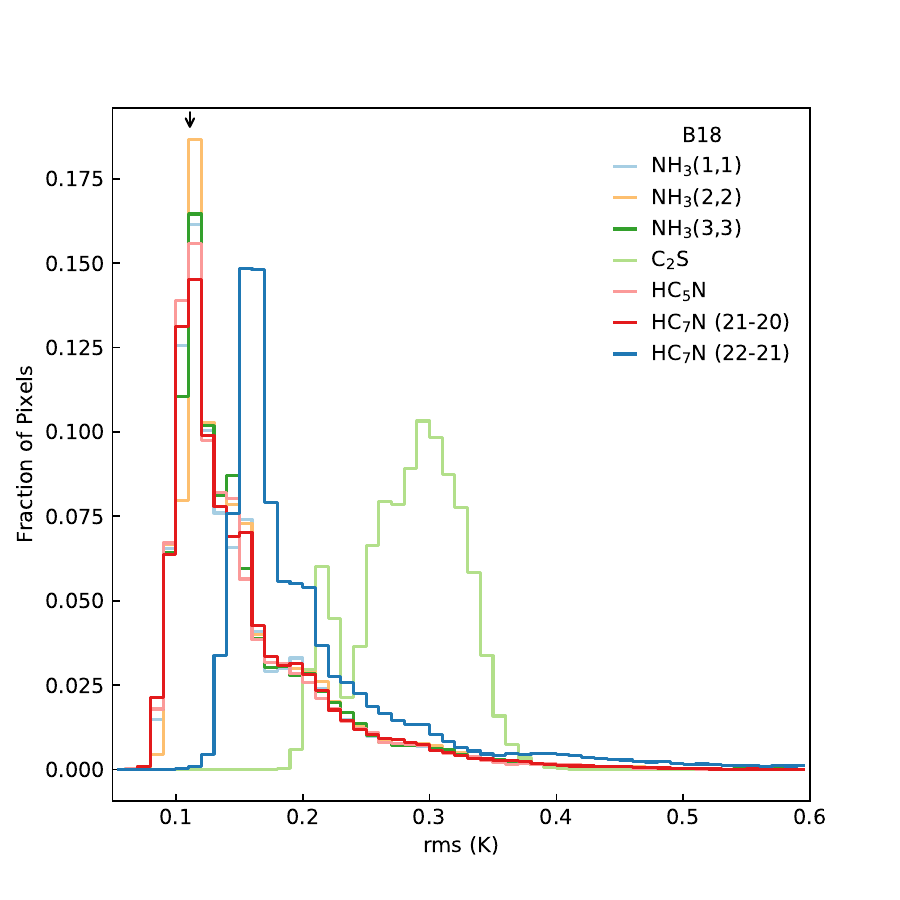}
\figsetgrpnote{B18 histogram of noise values per pixel for all observed lines. The black arrow shows the mode of the rms values for the NH3 (1,1) spectral window.}
\figsetgrpend

\figsetgrpstart
\figsetgrpnum{10.3}
\figsetgrptitle{B1E}
\figsetplot{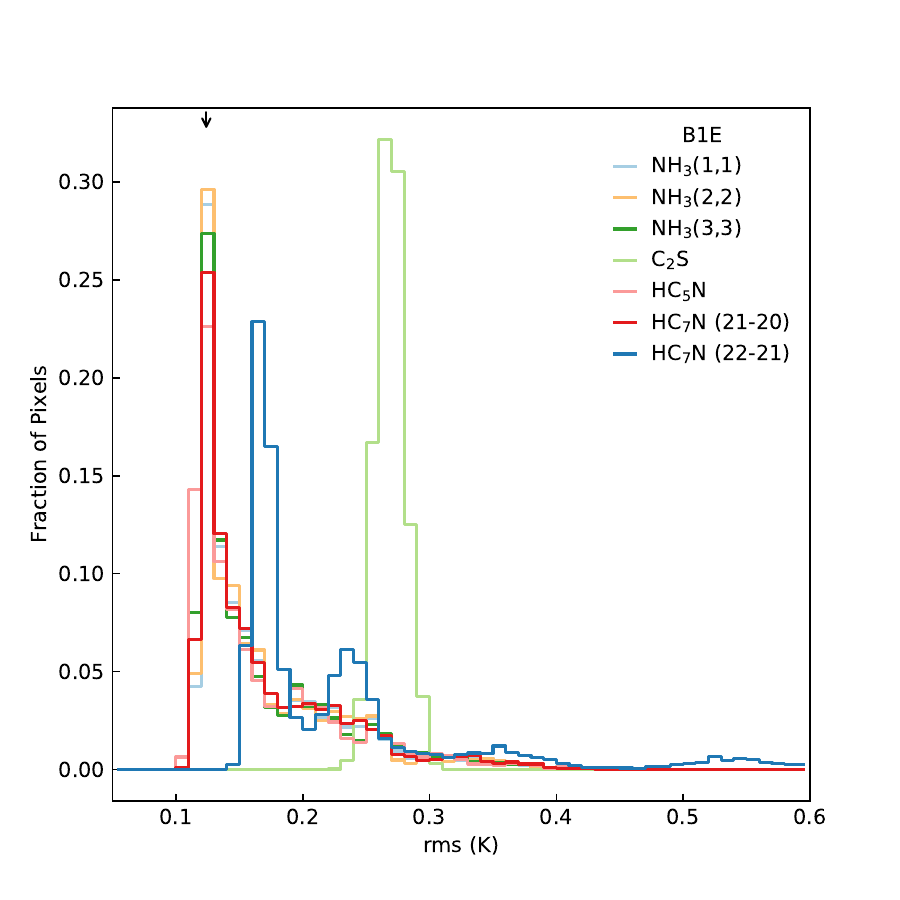}
\figsetgrpnote{B1E histogram of noise values per pixel for all observed lines. The black arrow shows the mode of the rms values for the NH3 (1,1) spectral window.}
\figsetgrpend

\figsetgrpstart
\figsetgrpnum{10.4}
\figsetgrptitle{B59}
\figsetplot{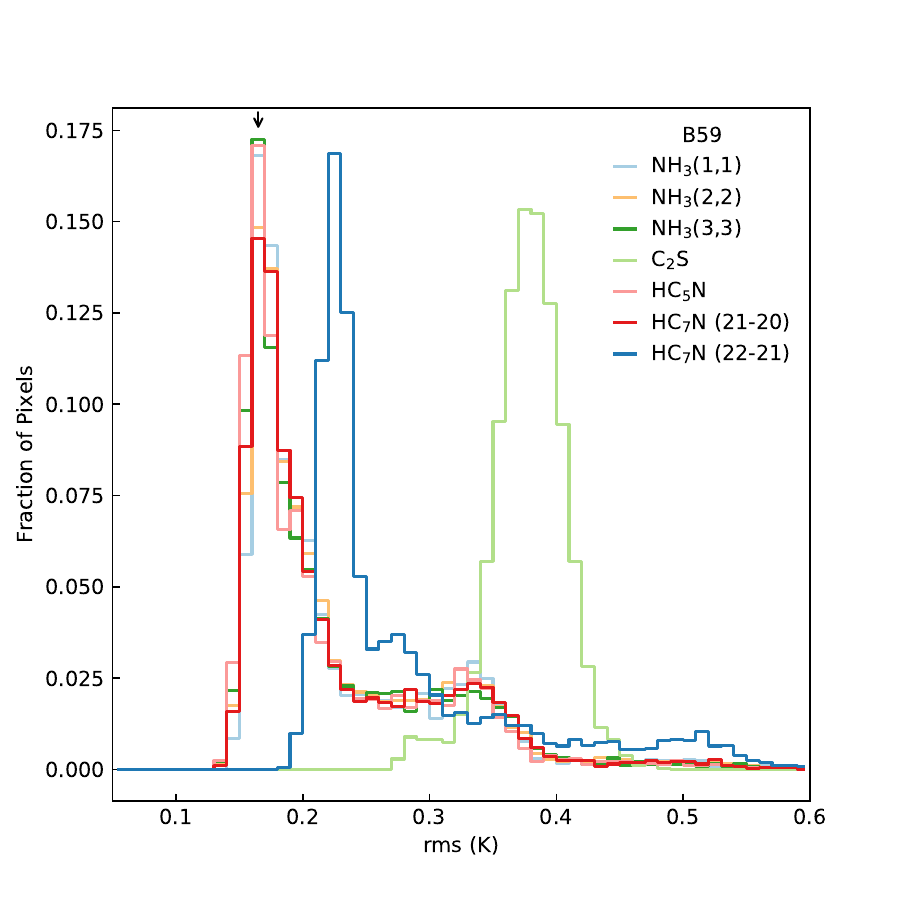}
\figsetgrpnote{B59 histogram of noise values per pixel for all observed lines. The black arrow shows the mode of the rms values for the NH3 (1,1) spectral window.}
\figsetgrpend

\figsetgrpstart
\figsetgrpnum{10.5}
\figsetgrptitle{L1228}
\figsetplot{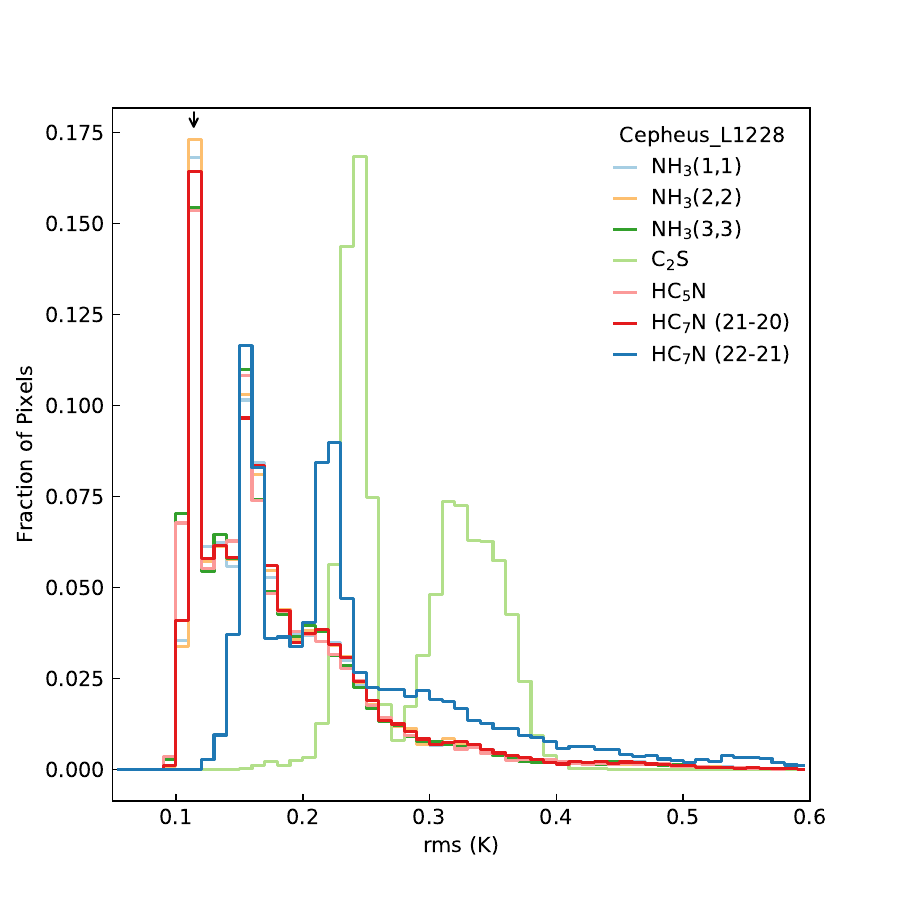}
\figsetgrpnote{L1228 histogram of noise values per pixel for all observed lines. The black arrow shows the mode of the rms values for the NH3 (1,1) spectral window.}
\figsetgrpend

\figsetgrpstart
\figsetgrpnum{10.6}
\figsetgrptitle{L1251}
\figsetplot{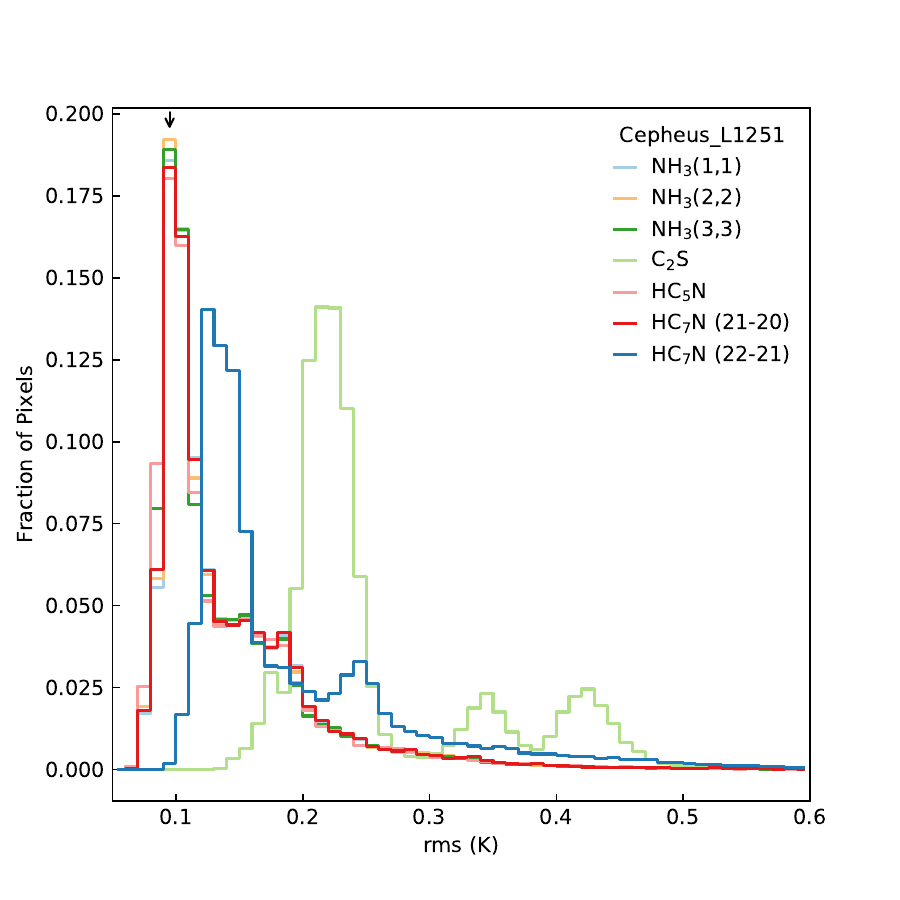}
\figsetgrpnote{L1251 histogram of noise values per pixel for all observed lines. The black arrow shows the mode of the rms values for the NH3 (1,1) spectral window.}
\figsetgrpend

\figsetgrpstart
\figsetgrpnum{10.7}
\figsetgrptitle{CrA East}
\figsetplot{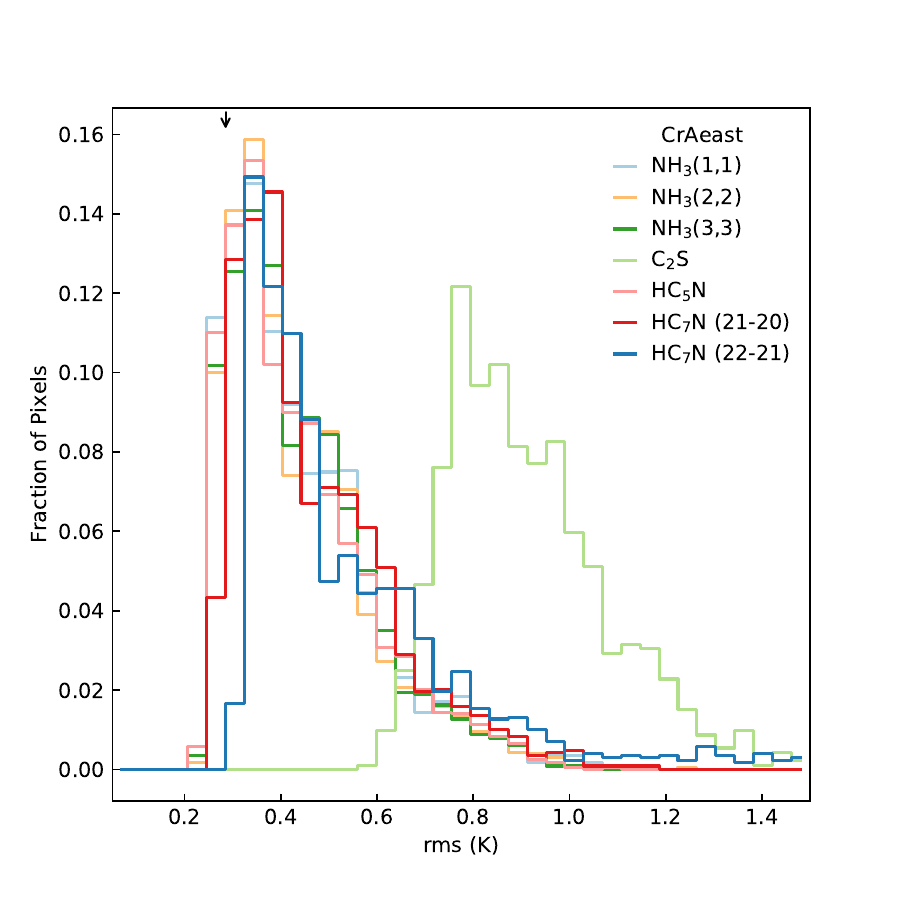}
\figsetgrpnote{CrA East histogram of noise values per pixel for all observed lines. The black arrow shows the mode of the rms values for the NH3 (1,1) spectral window.}
\figsetgrpend

\figsetgrpstart
\figsetgrpnum{10.8}
\figsetgrptitle{CrA West}
\figsetplot{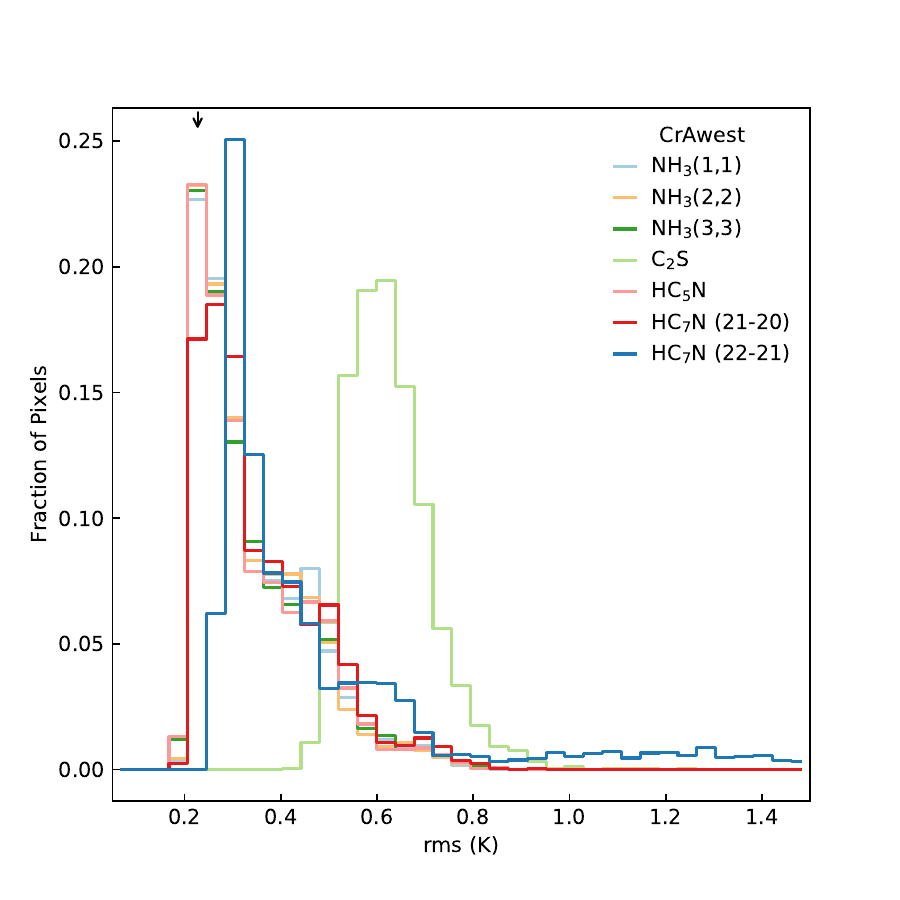}
\figsetgrpnote{CrA West histogram of noise values per pixel for all observed lines. The black arrow shows the mode of the rms values for the NH3 (1,1) spectral window.}
\figsetgrpend

\figsetgrpstart
\figsetgrpnum{10.9}
\figsetgrptitle{HC2}
\figsetplot{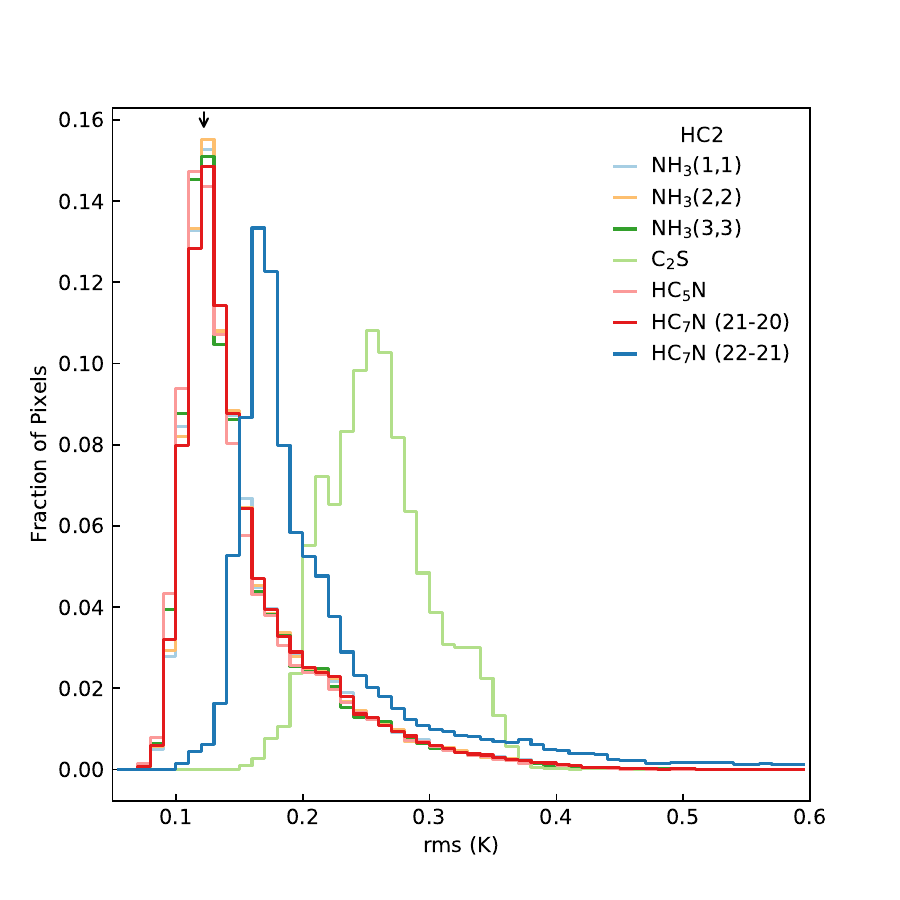}
\figsetgrpnote{HC2 histogram of noise values per pixel for all observed lines. The black arrow shows the mode of the rms values for the NH3 (1,1) spectral window.}
\figsetgrpend

\figsetgrpstart
\figsetgrpnum{10.10}
\figsetgrptitle{IC 348}
\figsetplot{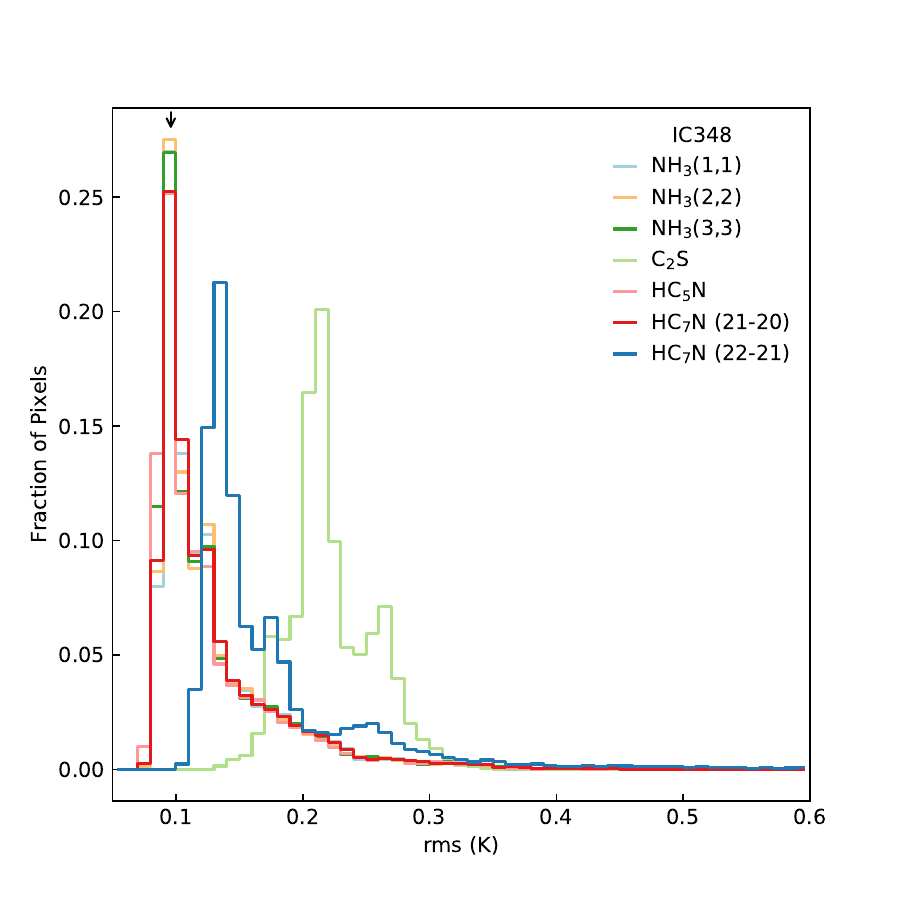}
\figsetgrpnote{IC 348 histogram of noise values per pixel for all observed lines. The black arrow shows the mode of the rms values for the NH3 (1,1) spectral window.}
\figsetgrpend

\figsetgrpstart
\figsetgrpnum{10.11}
\figsetgrptitle{IC 5146}
\figsetplot{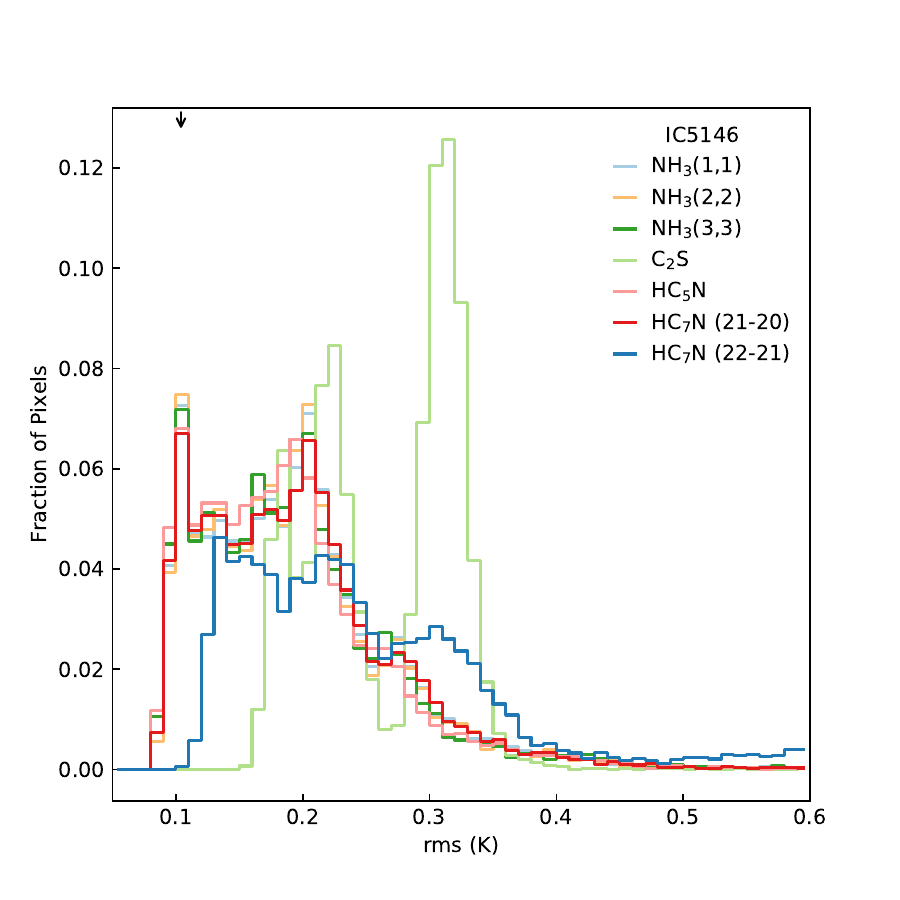}
\figsetgrpnote{IC 5146 histogram of noise values per pixel for all observed lines. The black arrow shows the mode of the rms values for the NH3 (1,1) spectral window.}
\figsetgrpend

\figsetend

\begin{figure*}
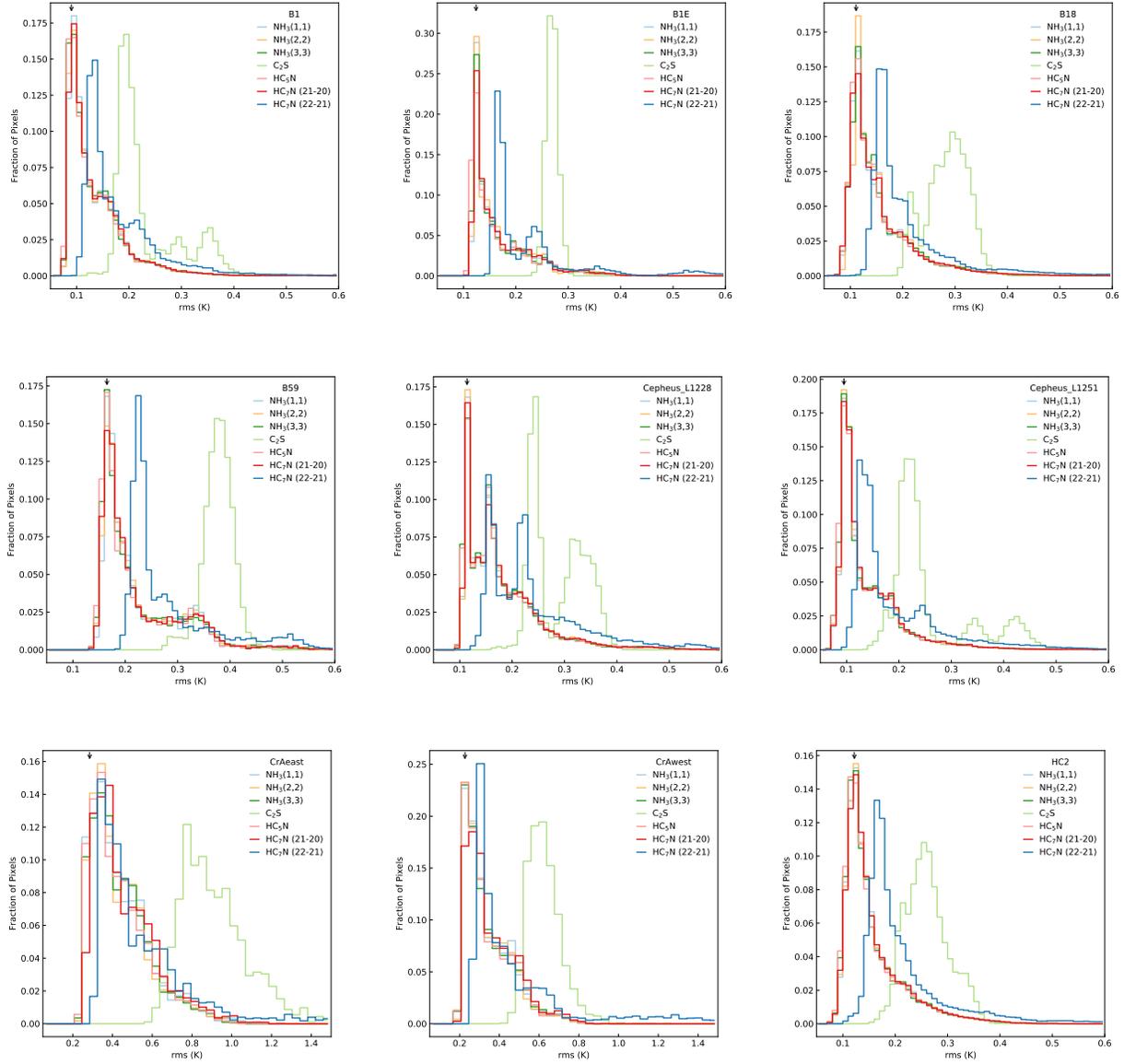

\epsscale{0.35}
\plotone{B1_all_all_rebase3_rms_hist}
\plotone{B1E_all_all_rebase3_rms_hist}
\plotone{B18_all_all_rebase3_rms_hist}
\plotone{B59_all_all_rebase3_rms_hist}
\plotone{Cepheus_L1228_all_all_rebase3_rms_hist}
\plotone{Cepheus_L1251_all_all_rebase3_rms_hist}
\plotone{CrAeast_all_all_rebase3_rms_hist}
\plotone{CrAwest_all_all_rebase3_rms_hist}
\plotone{HC2_all_all_rebase3_rms_hist}
\epsscale{1}
\caption{Histograms of noise values per pixel toward each of the GAS regions for all observed lines. The black arrow shows the mode of the rms values for the \amm (1,1) spectral window. The complete collection of histograms is available as a figure set in the online journal.
\label{fig:rms}}
\end{figure*}

\begin{deluxetable*}{lcccccccccccccc}
\tablecolumns{15}
\tablewidth{0pt}
\tablecaption{Noise properties of line maps by region \label{tab:noise}}
\tablehead{
\colhead{Region} & \multicolumn{2}{c}{\amm (1,1)} & \multicolumn{2}{c}{\amm (2,2)} & \multicolumn{2}{c}{\amm (3,3)} & \multicolumn{2}{c}{\ccs $2_1 - 1_0$} & 
\multicolumn{2}{c}{\hcfn 9-8} & \multicolumn{2}{c}{\hcsn 21-20} & \multicolumn{2}{c}{\hcsn 22-21}\\
\colhead{} & \colhead{mode} & \colhead{mad} & \colhead{mode} & \colhead{mad} & \colhead{mode} & \colhead{mad} & \colhead{mode} & \colhead{mad} & \colhead{mode} & \colhead{mad} & \colhead{mode} & \colhead{mad} & \colhead{mode} & \colhead{mad}
}
\startdata
B1 & 0.09 & 0.04 & 0.09 & 0.04 & 0.09 & 0.04 & 0.19 & 0.03 & 0.09 & 0.04 & 0.09 & 0.04 & 0.13 & 0.05 \\ 
B1E & 0.12 & 0.03 & 0.12 & 0.03 & 0.12 & 0.03 & 0.27 & 0.01 & 0.12 & 0.03 & 0.12 & 0.03 & 0.17 & 0.04 \\ 
B18 & 0.11 & 0.04 & 0.11 & 0.04 & 0.11 & 0.04 & 0.30 & 0.04 & 0.11 & 0.04 & 0.11 & 0.04 & 0.16 & 0.04 \\ 
B59 & 0.17 & 0.04 & 0.17 & 0.04 & 0.17 & 0.04 & 0.38 & 0.03 & 0.16 & 0.04 & 0.17 & 0.04 & 0.23 & 0.05 \\ 
Cepheus L1228 & 0.11 & 0.06 & 0.12 & 0.06 & 0.11 & 0.06 & 0.24 & 0.07 & 0.11 & 0.06 & 0.12 & 0.06 & 0.16 & 0.09 \\ 
Cepheus L1251 & 0.10 & 0.04 & 0.09 & 0.04 & 0.09 & 0.03 & 0.22 & 0.03 & 0.09 & 0.04 & 0.10 & 0.04 & 0.13 & 0.05 \\ 
CrAeast & 0.28 & 0.14 & 0.34 & 0.13 & 0.40 & 0.13 & 0.78 & 0.16 & 0.33 & 0.13 & 0.39 & 0.14 & 0.36 & 0.19 \\ 
CrAwest & 0.23 & 0.10 & 0.22 & 0.10 & 0.23 & 0.10 & 0.61 & 0.08 & 0.23 & 0.10 & 0.24 & 0.11 & 0.30 & 0.14 \\ 
HC2 & 0.12 & 0.04 & 0.12 & 0.03 & 0.12 & 0.03 & 0.26 & 0.04 & 0.12 & 0.03 & 0.12 & 0.04 & 0.17 & 0.04 \\ 
IC348 & 0.10 & 0.03 & 0.09 & 0.03 & 0.09 & 0.03 & 0.21 & 0.03 & 0.09 & 0.03 & 0.10 & 0.03 & 0.13 & 0.03 \\ 
IC5146 & 0.10 & 0.07 & 0.11 & 0.07 & 0.10 & 0.07 & 0.31 & 0.07 & 0.10 & 0.07 & 0.10 & 0.07 & 0.14 & 0.12 \\ 
L1448 & 0.12 & 0.02 & 0.12 & 0.02 & 0.11 & 0.02 & 0.27 & 0.02 & 0.11 & 0.02 & 0.12 & 0.02 & 0.16 & 0.03 \\ 
L1451 & 0.12 & 0.03 & 0.11 & 0.03 & 0.12 & 0.03 & 0.25 & 0.02 & 0.11 & 0.03 & 0.11 & 0.03 & 0.15 & 0.03 \\ 
L1455 & 0.10 & 0.02 & 0.10 & 0.02 & 0.10 & 0.02 & 0.23 & 0.03 & 0.10 & 0.02 & 0.10 & 0.02 & 0.15 & 0.04 \\ 
L1688 & 0.12 & 0.03 & 0.12 & 0.03 & 0.12 & 0.03 & 0.32 & 0.06 & 0.11 & 0.03 & 0.11 & 0.03 & 0.16 & 0.05 \\ 
L1689 & 0.13 & 0.04 & 0.13 & 0.04 & 0.13 & 0.04 & 0.30 & 0.03 & 0.13 & 0.04 & 0.14 & 0.04 & 0.18 & 0.04 \\ 
L1712 & 0.14 & 0.05 & 0.14 & 0.05 & 0.14 & 0.04 & 0.32 & 0.02 & 0.14 & 0.04 & 0.14 & 0.04 & 0.19 & 0.05 \\ 
NGC1333 & 0.11 & 0.01 & 0.11 & 0.01 & 0.10 & 0.02 & 0.27 & 0.02 & 0.11 & 0.01 & 0.10 & 0.01 & 0.15 & 0.03 \\ 
OrionA & 0.10 & 0.03 & 0.10 & 0.03 & 0.10 & 0.03 & 0.23 & 0.05 & 0.11 & 0.03 & 0.10 & 0.03 & 0.15 & 0.05 \\ 
OrionA S & 0.15 & 0.06 & 0.15 & 0.06 & 0.15 & 0.06 & 0.35 & 0.05 & 0.14 & 0.06 & 0.16 & 0.07 & 0.20 & 0.07 \\ 
OrionB NGC2023-2024 & 0.09 & 0.03 & 0.10 & 0.03 & 0.09 & 0.03 & 0.20 & 0.04 & 0.09 & 0.03 & 0.10 & 0.03 & 0.13 & 0.04 \\ 
OrionB NGC2068-2071 & 0.19 & 0.07 & 0.19 & 0.07 & 0.19 & 0.07 & 0.41 & 0.12 & 0.19 & 0.07 & 0.20 & 0.07 & 0.26 & 0.10 \\ 
Perseus & 0.18 & 0.07 & 0.17 & 0.07 & 0.17 & 0.07 & 0.45 & 0.06 & 0.17 & 0.07 & 0.18 & 0.07 & 0.24 & 0.07 \\ 
Pipe Core40 & 0.18 & 0.07 & 0.18 & 0.07 & 0.18 & 0.07 & 0.42 & 0.02 & 0.17 & 0.07 & 0.18 & 0.07 & 0.24 & 0.08 \\ 
Serpens Aquila & 0.13 & 0.03 & 0.13 & 0.03 & 0.13 & 0.03 & 0.29 & 0.05 & 0.13 & 0.03 & 0.27 & 0.11 & 0.16 & 0.03 \\ 
Serpens MWC297 & 0.18 & 0.06 & 0.18 & 0.06 & 0.18 & 0.06 & 0.46 & 0.05 & 0.19 & 0.06 & 0.20 & 0.07 & 0.24 & 0.07 \\ 
\enddata
\end{deluxetable*} 

We show in Fig.~\ref{fig:rms}  histograms of the rms noise per pixel toward each GAS region for each spectral window. 
The complete figure set (26 subplots) is available in the online journal. 
The Figure shows that the rms values are not Gaussian, with the majority of pixels having rms values around one value, but with a significant tail to greater rms. 
As mentioned earlier, this behavior is largely due to the non-uniform coverage of the maps from scanning with the hexagonal arrangement of seven beams with beam centers separated by $\approx$95\arcsec\ in the KFPA, but also results from varying weather conditions in the different map blocks. 
To best describe the noise properties of the majority of the data, we list in Table \ref{tab:noise} the mode (rather than the mean or median) and 
median absolute deviation (MAD, rather than the standard deviation) of the rms distribution for each region, for each line observed. 
To calculate the mode, we round the rms values to three decimal places.  
For most regions, the mode of the rms lies at roughly $\sim 0.1$~K for the \amm, \hcfn, and \hcsn\, 21--20 lines, 
with larger values for \ccs\, and \hcsn\, 22--21 as noted above. 
In general, median rms values tend to be larger than the mode by $\sim 10 - 20$~\%. 

\section{Column densities and abundances of \ce{C2S} and \ce{HC7N}}
\label{sec:app_col}

Here we provide similar measurements of the column densities of \ce{C2S} and \ce{HC7N}, and their abundances with respect to \amm  and \ce{H2}, as in Table \ref{tab:ccsResults} and \ref{tab:hc7nResults}, respectively. 
For \ce{C2S}, we list the median and (16$^{th}$, 84$^{th}$) values for each parameter, while we only provide the mean values for \ce{HC7N} due to its limited distribution. The column densities and abundances of \ce{HC7N} were calculated based on the $J = 21-20$ rotational transition only. 

\floattable 
\begin{deluxetable}{lccccc} 
\tabletypesize{\footnotesize} 
\tablecolumns{6} 
\tablewidth{0pt} 
\tablecaption{Regions with C$_2$S detections \label{tab:ccsResults}} 
\tablehead{ 
\colhead{Region} & \colhead{log $N$(C$_2$S)} & 
\colhead{log $N$(NH$_3$)} & 
\colhead{log $X$(C$_2$S/NH$_3$)} & 
\colhead{log $N$(H$_2$)} & 
\colhead{log $X$(C$_2$S/H$_2$)} } 
\startdata 
B1 & 13.00$^{0.16}_{-0.18}$ & 14.13$^{0.26}_{-0.28}$ & -1.11$^{0.25}_{-0.24}$ & 21.05$^{0.47}_{-0.33}$ & -9.11$^{0.28}_{-0.24}$ \\ 
B18 & 13.14$^{0.24}_{-0.16}$ & 14.00$^{0.23}_{-0.20}$ & -0.78$^{0.34}_{-0.33}$ & 21.18$^{0.29}_{-0.21}$ & -8.87$^{0.28}_{-0.26}$ \\ 
HC2 & 13.26$^{0.28}_{-0.25}$ & 13.95$^{0.30}_{-0.18}$ & -0.57$^{0.36}_{-0.41}$ & 21.49$^{0.21}_{-0.24}$ & -8.76$^{0.26}_{-0.24}$ \\ 
Serpens Aquila & 13.32$^{0.19}_{-0.19}$ & 14.06$^{0.30}_{-0.15}$ & -0.79$^{0.27}_{-0.41}$ & 21.56$^{0.21}_{-0.17}$ & -8.59$^{0.20}_{-0.21}$ \\ 
\enddata 
\tablecomments{Values given are the 50$^{th}$ percentile, along with the 16$^{th}$ and 84$^{th}$ percentiles of the parameter distributions for all columns. Column densities and abundances of \ccs relative to H$_2$ are calculated over all pixels with good line fits, while the \amm column densities and \ccs abundances relative to \amm are calculated in regions where both lines are detected. } 
\end{deluxetable} 

\floattable 
\begin{deluxetable}{lccccc} 
\tabletypesize{\footnotesize} 
\tablecolumns{6} 
\tablewidth{0pt} 
\tablecaption{Regions with HC$_7$N detections \label{tab:hc7nResults}} 
\tablehead{ 
\colhead{Region} & \colhead{log $N$(HC$_7$N)} & 
\colhead{log $N$(NH$_3$)} & 
\colhead{log $X$(HC$_7$N/NH$_3$)} & 
\colhead{log $N$(H$_2$)} & 
\colhead{log $X$(HC$_7$N/H$_2$)} } 
\startdata 
B1 & 13.31 &  \nodata &  \nodata & 20.84  & -7.54 \\ 
HC2 & 12.89 & 13.97 & -1.05 & 22.11  & -9.23 \\ 
\enddata 
\tablecomments{Values given are the mean values of the parameter distributions for all columns. Column densities and abundances of \hcsn relative to H$_2$ are calculated over all pixels with good line fits, based on the \hcsn $J = 21-20$ emission line. The \amm column densities and \hcsn abundances relative to \amm are calculated in regions where both lines are detected. We do not detect HC$_7$N in sufficiently large areas to provide percentiles. } 
\end{deluxetable} 

\section{Core catalog}
The core catalog is provided in Table~\ref{tab:core_cat}, with the full version available in the electronic version.
{The table include the region, the index number, the right ascension and declination coordinates, the core radius and mass, and 
the protostellar content as reported in \cite{Pandhi2023-GAS_Shape_VGrad}. 
In addition we derived the the mean LSR velocity, the mean observed velocity dispersion, the excitation and kinetic temperature, the \amm column density, 
the thermal and non-thermal velocity dispersions, the matched continuum ID, 
and the estimated viral parameters (see Eq.~\ref{eqn: Virial Mass} and \ref{eqn:sigma}).}

\begin{rotatetable*}
\begin{center}
\movetabledown=10mm
\begin{deluxetable*}{lllllrrrrrrlrrrl}
\tablecolumns{16}
\tabletypesize{\scriptsize}
\tablecaption{\amm core properties for selected regions\label{tab:core_cat}}
\tablehead{
  \colhead{Region} & \colhead{GAS} & \colhead{R.A.} &  \colhead{Dec.} & \colhead{\vlsr} & \colhead{\sigv} &
  \colhead{\tex} & \colhead{\tkin} &
  \colhead{$\log \namm$} & \colhead{$\sigma_{th}$} & \colhead{\signt} &
  \colhead{continuum id} & \colhead{$R$} &   \colhead{$M$} & \colhead{\avir} &          \colhead{Type} \\
  \colhead{} & \colhead{\tablenotemark{a}} & \colhead{(J2000)} & \colhead{(J2000)} & \colhead{(\kms)} & \colhead{(\kms)} &   \colhead{(K)} & \colhead{(K)} & \colhead{(\sqcm)}  & \colhead{(\kms)} & \colhead{(\kms)} &
  \colhead{} & \colhead{(pc)} & \colhead{(\msun)} & \colhead{} & \colhead{} 
}
\startdata
            B1 &       3 &   3:32:16.5 &   30:49:32.0 &      6.80 (0.11) &  0.21 (0.05) &   4.75 (1.01) &    11.05 (0.82) &             14.1 (0.2) &          0.20 &          0.20 &        033217.8+304948 & 0.01 &   1.5 &           0.6 &  protostellar \\
            B1 &       5 &   3:32:19.5 &   30:51:36.0 &      6.56 (0.07) &  0.20 (0.07) &   3.60 (0.30) &    11.30 (0.00) &             13.9 (0.0) &          0.20 &          0.19 &        033218.9+305148 & 0.03 &   0.1 &          17.4 &    prestellar \\
            B1 &       4 &   3:32:32.6 &   30:50:26.9 &      6.57 (0.08) &  0.24 (0.16) &   3.70 (0.51) &    10.71 (0.54) &             13.9 (0.1) &          0.19 &          0.22 &        033232.0+305030 & 0.03 &   0.9 &           3.9 &    prestellar \\
            B1 &      11 &   3:32:34.0 &   30:56:29.4 &      6.54 (0.05) &  0.38 (0.08) &   3.19 (0.13) &    11.82 (0.00) &             14.2 (0.1) &          0.20 &          0.38 &        033233.3+305627 & 0.04 &   0.2 &          44.3 &    prestellar \\
            B1 &       6 &   3:32:35.6 &   30:52:56.9 &      6.84 (0.13) &  0.17 (0.10) &   3.94 (0.70) &     9.99 (1.32) &             13.7 (0.1) &          0.19 &          0.15 &        033236.7+305306 & 0.04 &   0.4 &          10.0 &    prestellar \\
            B1 &      27 &   3:32:44.3 &   30:59:59.1 &      6.72 (0.11) &  0.20 (0.07) &   4.99 (0.81) &    10.26 (0.73) &             14.1 (0.2) &          0.19 &          0.18 &        033243.7+305948 & 0.05 &   3.1 &           1.5 &    prestellar \\
            B1 &      58 &   3:33:01.9 &   31:20:53.3 &      6.64 (0.03) &  0.28 (0.04) &   4.07 (0.60) &    10.84 (2.20) &             14.0 (0.1) &          0.19 &          0.27 &        033300.8+312047 & 0.05 &   1.4 &           4.4 &    prestellar \\
            B1 &      34 &   3:33:03.1 &   31:04:33.0 &      6.62 (0.05) &  0.16 (0.03) &   5.95 (0.52) &     9.92 (0.46) &             14.2 (0.2) &          0.19 &          0.15 &        033302.5+310432 & 0.02 &   0.8 &           1.8 &    prestellar \\
            B1 &      40 &   3:33:05.4 &   31:06:30.5 &      6.57 (0.02) &  0.16 (0.01) &   5.59 (0.24) &    10.09 (0.35) &             14.2 (0.1) &          0.19 &          0.14 &        033305.0+310640 & 0.02 &   1.6 &           0.9 &    prestellar \\
            B1 &      43 &   3:33:16.5 &   31:06:59.5 &      6.33 (0.08) &  0.25 (0.01) &   7.78 (0.31) &    11.42 (0.20) &             14.7 (0.0) &          0.20 &          0.24 &        033316.4+310652 & 0.01 &   0.8 &           1.4 &  protostellar \\
            B1 &      45 &   3:33:18.0 &   31:09:19.8 &      6.32 (0.19) &  0.31 (0.04) &   7.00 (0.59) &    11.78 (0.72) &             14.5 (0.1) &          0.20 &          0.30 &        033317.7+310932 & 0.01 &   2.7 &           0.6 &  protostellar 
\enddata
\tablenotetext{a}{GAS ID from dendrogram analysis as listed in \cite{Pandhi2023-GAS_Shape_VGrad}. 
Note that GAS core IDs are unique only within regions.}
\tablecomments{\amm\ core catalog cross-matched with submillimeter continuum core catalogs, first presented by \cite{Pandhi2023-GAS_Shape_VGrad}. 
Here we provide mean and standard deviation values for \amm\ fit parameters within the \amm\ core contours as determined from the dendrogram analysis. Continuum core parameters, and core type, are taken from the continuum catalogs cited in \citeauthor{Pandhi2023-GAS_Shape_VGrad}, or determined directly therein. }
\end{deluxetable*}
\end{center}
\end{rotatetable*} 

{\section{Comparison Temperature and Velocity Dispersion\label{sec:Tk_sigmav}}
Figure~\ref{fig:violins} shows a correlation between the velocity dispersion, \sigv, and the gas kinetic temperature, \tkin, across different regions using violin plots. 
Figure~\ref{fig:Tk_dv} further demonstrates that the typical velocity dispersion in the different regions cannot be explained solely by an increase in \tkin. 
The data span from subsonic levels of non-thermal velocity dispersion to supersonic values, reaching up to $\mathcal{M}_s=2$.
This is similar to the strong correlation found by \cite{Friesen2024-Serpens_South_EVLA} in Serpens-South.

\begin{figure}[h]
\includegraphics[width=\columnwidth]{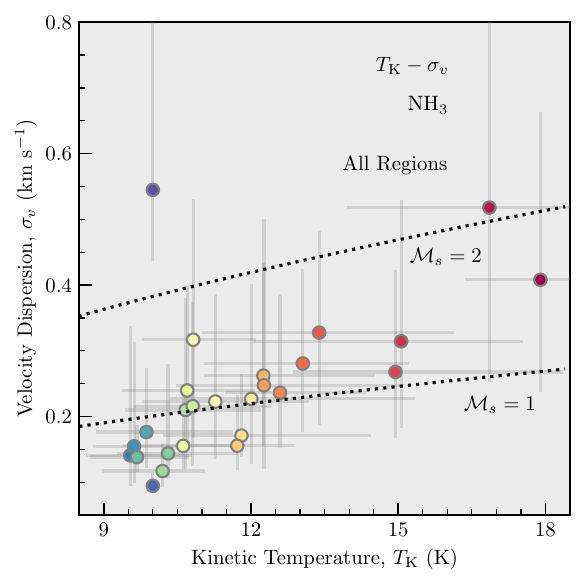}
\caption{Comparison of typical velocity dispersion and kinetic temperature for all regions.  
Symbols show the 50$^{\mathrm{th}}$, 16$^{\mathrm{th}}$, and 84$^{\mathrm{th}}$ percentiles of the values for each region, with asymmetric uncertainties. 
The color of the symbols corresponds to the mean \tkin, consistent with Fig.~\ref{fig:violins}. 
The dotted lines indicate the expected velocity dispersions for sonic Mach numbers of $\mathcal{M}_s=1$ and $\mathcal{M}_s=2$.
\label{fig:Tk_dv}}
\end{figure}
}

\bibliography{biblio}
\bibliographystyle{aasjournalv7}

\end{document}